\documentclass[10pt,twocolumn]{IEEEtran}
\usepackage{graphicx} % for pdf, bitmapped graphics files
\usepackage{epsfig} % for postscript graphics files
\usepackage{mathptmx} % assumes new font selection scheme installed
\usepackage{times} % assumes new font selection scheme installed
\usepackage{amsmath} % assumes amsmath package installed
\usepackage{amssymb}  % assumes amsmath package installed
\usepackage{color}
\usepackage{authblk}
\usepackage[caption=false,font=footnotesize]{subfig}
\usepackage{setspace}
\usepackage{algorithm}
\usepackage{algorithmic}
\DeclareMathOperator*{\argmax}{argmax}

\newtheorem{theorem}{Theorem}
\newtheorem{lemma}{Lemma}

\newtheorem{definition}{Definition}

\newcommand{\paren}[1]{\left(#1\right)}
\newcommand{\sqparen}[1]{\left[#1\right]}
\newcommand{\brparen}[1]{\left\{#1\right\}}
\newcommand{\field}[1]{\ensuremath{\mathbb{#1}}}
\newcommand{\N}{\ensuremath{\field{N}}} % natural numbers
\newcommand{\R}{\ensuremath{\field{R}}} % real numbers
\newcommand{\Rp}{\ensuremath{\R_+}} % positive real numbers
\newcommand{\vecbold}[1]{\ensuremath{\boldsymbol{#1}}}

\newcommand{\E}[1]{{\mathbb{E}}\left[{#1}\right]}

\newcommand{\Prob}[1]{{\mathbb{P}}\left({#1}\right)}

\newcommand{\Qv}{{\boldsymbol{Q}}}

\renewcommand\boldsymbol[1]{\pmb{#1}}

\begin{document}
%\onehalfspacing
\title{{ Dynamic Control of Interference Limited Underlay D2D Network}}%{\thanks{This material is based upon work supported by the National
%Science Foundation under Grants CNS-0831919, CCF-0916664, CAREER-1054738.}}\thanks{Portions of this work were presented at Asilomar Conference on Signals, Systems, and Computers (Asilomar '10), Pacific Grove, CA.}}
% author names and affiliations
% use a multiple column layout for up to three different
% affiliations
%\author{\IEEEauthorblockN{C.~Emre Koksal}
%\IEEEauthorblockA{Department of Electrical and Computer Engineering,\\
%The Ohio State University\\
%Columbus, OH \\
%Email: koksal@ece.osu.edu} \and \IEEEauthorblockN{Ozgur Ercetin and
%Yunus Sarikaya}
%\IEEEauthorblockA{Faculty of Engineering and Natural Sciences,\\
%Sabanci University,\\
%Istanbul, TR.\\
%Email: \{oercetin,ysarikaya\}@sabanciuniv.edu}}

\author{{ Yunus Sarikaya, Hazer Inaltekin, Tansu Alpcan and Jamie Evans}%
\thanks{Y. Sarikaya, J. Evans (\{yunus.sarikaya, jamie.evans\}@monash.edu) and T. Alpcan (tansu.alpcan@unimelb.edu.au) are with the Department of Electrical and Computer 
Systems Engineering at Monash University, Australia}  \thanks{H. Inaltekin (hazeri@antalya.edu.tr)  is with the Department of Electronics Engineering, Antalya International University, Turkey.} }%
%\thanks{C.~E. Koksal (koksal@ece.osu.edu) is with the Department of Electrical and Computer Engineering at The Ohio State University, Columbus, OH.}  \thanks{O. Ercetin (email: oercetin@sabanciuniv.edu) and  Y.Sarikaya (email: sarikaya@su.sabanciuniv.edu) are with the Department of Electronics Engineering, Faculty of Engineering and Natural Sciences, Sabanci University, 34956 Istanbul, Turkey.}}

\maketitle

\begin{abstract}

Device-to-Device (D2D) communication appears as a key communication paradigm to realizing the vision of Internet of Things (IoT) into reality by supporting heterogeneous objects interconnection in a large scale network. These devices may be many types of objects with embedded intelligence and communication capabilities, e.g., smart phones, cars, or home appliances. The issue in in this type of communication is the interference to cellular communication caused by D2D communication. Thus, proper power control and resource allocation should be
coordinated in D2D network to prevent excessive interference and drastic decrease in the throughput of the cellular system. In
this paper, we consider the problem of cross-layer resource allocation
in time-varying cellular wireless networks with D2D communication and incorporate
average interference to cellular system as a quality-of-service constraint. Specifically, each D2D pair in the network injects packets to its queue, at rates chosen in order to maximize a global
utility function, subject to network stability and interference constraints. The interference constraint enforces an arbitrarily low interference to the cellular system caused by D2D communication.
We first obtain the stability region for the multiuser systems assuming
that the nodes have full channel state information (CSI) of their
neighbors. Then, we provide a joint flow control and scheduling scheme, which is proven to achieve a utility arbitrarily close to the maximum
achievable utility. Finally, we address the consequences
of practical implementation issue such as distributed scheduling by a designing algorithm, which is capable of taking advantage of diversity gain introduced by fading channels. We demonstrate the
efficacy of our policies by numerical studies under various network
conditions.
%Numerical experiments are performed to verify
%the analytical results and to show the efficacy of the dynamic
%control algorithm.
\end{abstract}
\section{Introduction}

The explosive increase of user demands leads to emergence of new data intensive applications to accommodate this increasing demand of users. One of the examples are 4G cellular technologies (WiMAX and LTE-A), which have efficient physical and MAC layer performance, but are still lagging behind extremely fast-increasing users' data demand. Therefore, researches are seeking for new paradigm in the context of 5G technologies to increase the efficiency of wireless network.

While the conventional cellular architecture consists of connections from base stations to user equipment, 5G systems may well rely upon a two-tier architecture consisting of a macrocell tier for base station to device communication, and a second device tier for device to device (D2D) communications.  Such architectures are a hybrid of conventional cellular and adhoc designs. D2D communication in cellular networks is defined as direct communication between two mobile users without traversing the access point (AP) or first tier network.  Hence directly communicating with nearby wireless device  can highly increase the efficiency of the network by reusing cellular spectrum resources. This concept not only improves the efficiency of spectrum usage, but also has a great potential for enhancing the network performance expressed in terms of capacity, coverage, energy efficiency and end-to-end delays \cite{Lei12,Fodor12}. For the above benefits, D2D communication is one of such technologies that appears to be a promising component in next generation cellular network.

In the literature, various paradigms are introduced based on the spectrum in which D2D communication occurs. These paradigms are classified as overlay and underlay inband and outband communications. The paradigm of interest in this paper is underlay inband communications, in which both D2D and cellular devices use the same spectrum to transmit their data. Operating in the same licensed band, devices
will inevitably impact macrocell users by causing interference. To ensure minimal impact on the performance of existing APs, a two-tier network needs to be designed with smart interference management strategies and appropriate resource allocation schemes. Furthermore, users in today's cellular networks use real-time high data rate services like video sharing, gaming or proximity-aware social network. Those applications require end-to-end delay of incoming packets to be finite. Hence, the stability of user queues, i.e., the network stability, guaranteeing the delay to be finite, is another important design parameter, which should be taken into account. 

Considering those design parameters, we combine a variety of basic networking
mechanisms such as flow control and scheduling in the context of D2D underlaying cellular network. For that purpose, we
model the entire problem as that of a network utility maximization,
in which interference caused by D2D pairs to cellular BS is incorporated as an additional
constraint and develop the associated dynamic flow control,
power control, and scheduling mechanisms.

Scheduling in wireless networks is a prominent and challenging
problem that attracted significant interest from the
networking community. The challenge arises from the fact
that the capacity of wireless channel is time-varying nature of wireless channels. Optimal scheduling in wireless networks
has been extensively studied in the literature under various
assumptions \cite{tassiulas,shroff,urgaonkar,jaramillo}. Starting with the seminal work of
Tassiulas and Ephremides \cite{tassiulas}, where throughput optimality of
back-pressure algorithm is proven, policies opportunistically
exploiting the time-varying nature of the wireless channel
to schedule users are shown to be at least as good as static
policies \cite{shroff}. In principle, these opportunistic policies schedule
the user with the favorable channel condition to increase the
overall performance of the system. However, without imposing
individual performance guarantees for each user in the system,
this type of scheduling may result in unfair sharing of resources
and may lead to starvation of some users, for example, those
far away from the base station in a cellular network. Hence,
in order to address fairness issues, scheduling problem was
investigated jointly with the network utility maximization
problem \cite{kelly, Kar}, and the stochastic network optimization
framework that guarantees the network stability and provides fairness, was developed \cite{Georgiadis}.

In our model, we consider a D2D network in which communication takes place regardless of cellular system as long as the interference caused by them to AP is kept low to guarantee a required level of Quality-of-Service (QoS) on the cellular network. We call such network \textit{interference-aware} D2D network, In such setting, we use stochastic optimization framework to address the basic wireless network control problem
in order to develop a cross-layer resource allocation solution that
will incorporate interference metric to the AP as a QoS
metric. Our main contributions are summarized as follows:

\begin{enumerate}

\item[{\bf (a)}] We evaluate the \textit{ stability region} of interference-aware D2D network, defined as the set of rates achievable by any flow control and link-scheduling scheme. Then, we compare it with the stability region where there is no
interference constraint, and quantify the rate loss due to interference-aware operation of D2D network.

\item[{\bf (b)}] We formulate the
resource allocation problem in interference-aware D2D network as a network utility maximization (NUM) problem, in which the optimal scheduling of D2D communication links allocation is
implemented in the physical layer, while  the flow control is realized in the transport layer.

\item[{\bf(c)}] Although there is no standard for D2D communications, D2D communications in cellular network are expected to be overseen/controlled by a central entity like evolved Node N (eNB). Thus, our first focus is to design a centralized algorithm. We propose a dynamic control algorithm, which is a joint scheduling and flow control
scheme as the solution of the NUM problem, and  we prove that this scheme achieves a utility arbitrarily close to the maximum
achievable utility. Note that flow control algorithm moves the rate vector to the pareto boundary of the stability region based on defined utilities for the users. For example, logarithmic utility function moves the rates to the point where proportional fairness is satisfied.

\item[{\bf (d)} ]  We investigate important practical
limitation, which is the unavailability of a centralized scheduler.
Hence, we design a distributed scheduling algorithm called \textit{ channel-aware} distributed scheduler, where the devices decide to transmit or not based on their local information (i.e., based on their channel conditions and queue backlogs), and the proposed algorithm is channel-aware in the sense that they have capability of taking into account wireless channel variations. Then, we obtain performance bounds on the algorithms since the optimality
can no longer be guaranteed due to availability of limited information.
advantage of diversity gain imposed by fading channels. Then, we obtain performance bounds on the algorithms since the optimality
can no longer be guaranteed due to availability of limited information.
%We combine a variety of strategies developed in the context of
%information theoretic secrecy with basic networking mechanisms in
%the cognitive radio network setting such as flow control and
%bandwidth and power allocations. For that purpose, 
\item[{\bf (e)} ] We demonstrate via simulations the performance of proposed algorithms comparing with other well-known distributed schedulers and verify the analytical results.
%\item[{\bf (c)} ] We show that the proposed optimal scheme has high
%complexity in terms of computational overhead.  Thus, we design a
%sub-optimal solution based on the idea that decoupling the part of
%the problem which imposes high complexity, and solving this part
%offline.

%\item[{\bf (c)} ]  We obtain a closed form solution of optimal jamming powers when the instantaneous channel state of the eavesdropper channel is not available.
%Such a closed form solution is non-existent in the literature. In the
%literature, works either employ search algorithms to find the
%optimal solution or make assumptions on the probability of secrecy
%outages \cite{M_Ghaderi}, \cite{J_Huang}, which leads to high
%complexity in terms of computational overheard or a sub-optimal
%solution.
\end{enumerate}

%We note that the proposed control algorithm is centralized.  

%D2D users may act autonomously only when the cellular infrastructure is unavailable.  

%throughput under the constraint of primary user's queue stability.

\section{Related Works}

Our results in this paper are related to optimization in D2D communication in underlaying cellular network. Hence, we will only mention the papers which are most relevant to ours.

The authors of \cite{Yu09} and \cite{Yu11} consider a single cell scenario where a cellular user and two D2D users share the same spectrum, where the BS controls the transmit power and spectrum of the D2D communication link. The objective is to optimize the sum rate with energy/power constraint under non-orthogonal and orthogonal sharing mode. The authors show analytically that an optimal solution can be given either in closed form or can be chosen from a set. Then, the researchers turn their eyes on the efficient resource allocation with multiple D2D pairs. 
The
work in \cite{Janis09} controls the interference of the D2D links to
the cellular users by limiting the maximum transmit power of
the D2D users. In
\cite{Min11}, authors employ the interference-limited area knowledge
for D2D receivers to maximize the network capacity with
multiuser MIMO, and any cellular users in the vicinity of
the interference-limited area is not scheduled. The optimal power allocation problem for D2D communication pairs are analyzed in \cite{Xiao11, Jung12, Hakola10, Belleschi11}. The authors in \cite{Xiao11} and \cite{Hakola10} showed that the problem of optimal power allocation and mode selection problem are rather involved and they propose heuristic approaches to solve the problem. The authors of \cite{Jung12} propose a method which applies exhaustive search  to find power efficiency, which is a function of transmission rate and power consumption of the users.

The works in \cite{Feng13, Zhang13, Su13,Han12,Le12} focus on the improving the system performance while maintaining certain QoS constraint. The authors in \cite{Feng13} consider a resource allocation problem to maximize the overall network throughput while guaranteeing QoS requirements for both D2D and regular users. A maximum weight bi partite matching based scheme is developed to select a suitable D2D communication pair. \cite{Zhang13} propose a graph-based resource allocation method, where they formulate the optimal resource allocation as a non-linear problem. Since the problem is NP-hard, they propose a suboptimal graph-based approach, which accounts for interference and capacity of the network. In their proposed graph, each vertex represents a link (D2D or cellular) and each edge connecting two vertices shows the potential interference the two links. \cite{Su13} formulate the problem of maximizing the system throughput with minimum data rate requirements, and they use the particle swarm optimization framework to obtain the solution. In \cite{Le12}, the formulate the problem as integer programming problem. which is hard to solve due to being NP-hard. Hence, they propose a sub-optimal solution which is based on obtaining the elements of optimization problem in different phases.

On the other hand, the problems when the cellular infrastructure is missing, i.e., distributed algorithms where D2D users act autonomously, are mostly analyzed in a game-theoretical framework. In \cite{Wang14}, the proposed approach optimally selects
the most beneficial source devices by analyzing the interactions
between the base station's rewarding strategies
and the devices' transmission
power using a Stackelberg game model. In addition to pricing model, \cite{Xu13} propose auction-based game theoretic models, where D2D communication pairs should bid for the channel to transmit their information. Other approaches propose a coalition formation game
based scheme \cite{Li14}, \cite{Chen14}, where cellular users and D2D communications pairs who want to communicate in the same spectrum, form a coalition. Most game-theoretical
solutions impose large overhead due to the need for heavy information exchange in terms of
bids, or prices and demands.

The solutions of these approaches and their derived sub-optimal heuristic can indeed improve the system performance with QoS constraints. However, they do not seem to be a good candidate for time-stringent application with limited computational capacity. Nonetheless, the authors  of \cite{Yu09} and \cite{Yu11} derived the closed-form solution, which reduces the complexity. But they only consider a scenario with a cellular user and D2D communication pair, which is not practical in reality. The algorithm derived in this paper, which is shown to be close to optimal solution, is a simple index policy, i.e., the centralized entity performs simple algebraic manipulations to obtain the solution in each time instant. Furthermore, we consider a general network model which contains arbitraty number of D2D communication pairs.
{\allowdisplaybreaks\section{System Model, Scheduling Policy and D2D Network Stability}
\label{sec:model}

In this section, we will introduce the details of our system model and the definitions of the main concepts that are used throughout the paper in relation to this model. 
\begin{figure}
\centerline{ \includegraphics[width=3.0in]{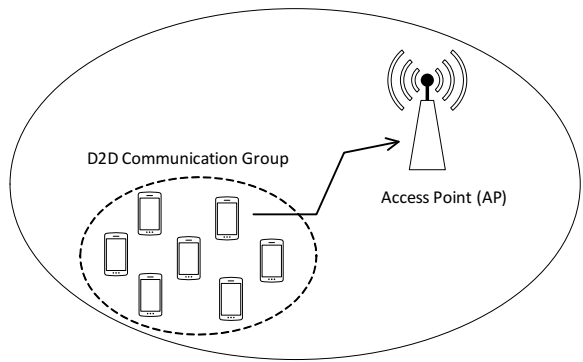}
} \caption{Network model consisting of $N$ D2D pairs that share a common frequency band with an access point.}
\label{fig:d2dnetwork}
\end{figure}

\subsection{System Model} 

Our primary aim in this paper is to characterize the maximum rates of communication that can be {\em stably} supported by a D2D network and to discover the cross-layer {\em centralized} and/or {\em distributed} control mechanisms (i.e., flow control and scheduling) that can achieve these rates with provable performance guarantees.  To this end, we consider a group of communication devices forming $N$ distinct D2D communication pairs and  sharing the same frequency band with an access point (AP), as shown in Fig. \ref{fig:d2dnetwork}.  The devices are in close proximity of each other so that they can reach to their intended receivers in a single hop, but they also cause excessive interference to each other when two or more pairs are active at the same time. This leads to a fully connected interference graph topology with collision model for the D2D network in question. %, i.e., bits over a link can be decoded successfully only when there is {\em one} active D2D %pair.          

%We consider an underlay radio network in which $N$ distinct D2D pairs share a frequency  band with the access point (AP) as shown in Fig. \ref{fig:d2dnetwork}. Each pair has data to be transmitted and its transmissions interfere with the signal
%reception at the AP. We assume fully connected interference graph with collision model, i.e., if multiple links are active at the same time instant, packet collisions take place and no data is transmitted over any link.

The devices operate in slotted time with slot indices represented by $t \in \N$. The link quality between the device pairs varies over time according to the {\em block} fading model, in which the channel gain is constant over a time slot and changes from one slot to another independently according to a common fading distribution. We use $h_i(t)$, $i = 1, \ldots, N$, to represent the {\em direct} channel gain between the transmitter and receiver of the $i$th D2D pair. These direct channel gains are independent and identically distributed (iid) over users as well as over time. Operating in the same frequency band, the devices also cause interference to the AP in Fig. \ref{fig:d2dnetwork}, and we denote the {\em interference} channel gain between the transmitter of the $i$th D2D pair and the AP by $g_i(t)$, $i = 1, \ldots, N$. Again, interference channel gains obey to the iid block fading model (possibly with a different distribution than that of the direct channel gains) as described above. For notational simplicity, we often use  the vector notation $\vecbold{h}(t) = \sqparen{h_1(t), \ldots, h_N(t)}$ and $\vecbold{g}(t) = \sqparen{g_1(t), \ldots, g_N(t)}$ to denote the channel gains more compactly.  

For the sake of comprehending the interplay between the scheduling decisions at the MAC layer and the flow control decisions at the transport layer better, it is assumed that no power control is exercised at the physical layer of the D2D network and all devices transmit at a constant power level $P$ over all time slots.  This assumption will help us to distill the effect of physical layer parameters on the interactions of the upper layer scheduling and flow control protocols, which is the main focus of the current paper.  In this setting, an important quantity of interest that determines the D2D network performance is the rates (measured in units of bits/slot) offered over a communication link during time slot $t$. We assume that these communication rates are described by the functions $R_i(t)$ (as functions of transmission power levels and channel gains) for $i=1, \ldots, N$. Even though we do not assume any specific functional form for $R_i(t)$, which is usually determined by the coding and communication technologies embedded in the transceiver circuits of the devices,\footnote{For example, if the Shannon capacity formula is used to quantify the communication rates for the $i$th D2D pair, $R_i(t)$ can be given as $R_i(t) = \log\paren{1 + \frac{Ph_i(t)}{N_0}}$, where $N_0$ represents the total noise plus interference power degrading transmissions over the $i$th link.  Indeed, these communication rates are achievable by using Gaussian codebooks when slot durations are large enough \cite{Tse98}.} we require that $R_i(t)$ has a bounded second moment, i.e., $\E{R_i(t)^2} \leq R_{max}^2$ for all $t \in \N$. The significance of the rate function $R_i(t)$ in our analysis is that it will determine the {\em service} rates of the network layer queues maintained at the devices.       

%Communication links vary over time according to independent and identically distributed (iid) block fading model, in which the channel gain is constant over a time slot and it is changing independently from slot to slot. We represent the direct channel gain in slot $t$ for the $i$th D2D link by $h_i(t)$, i.e.,  $h_i(t)$ is the channel gain between the transmitter of pair $i$ and the receiver of that pair. Similarly, the interference channel gain for the $i$th D2D communication link in slot $t$ is represented by $g_i(t)$, i.e., $g_i(t)$ is the channel gain between the transmitter of the $i$th link and the AP. We define the vector of channel gains by $\boldsymbol{h}(t) = [h_1(t), \ldots, h_N(t)]$ and $\boldsymbol{g}(t) = [g_1(t), \ldots, g_N(t)]$. We also denote the data rate of D2D pair $i$ in slot $t$ by $R_i(t)$. We assume that the transmit power of devices to be constant, identical to $P$ over all time slots $t$, i.e., $P_i(t) = P_j(t) = P, \ \forall j \neq i$. Based on coding/communication schemes, the channel rate is primarily determinant by the power level and channel state \footnote{ The maximum achievable rate defined as Shannon capacity is given as: $R_i(t) = \log(1 + Ph_i(t))$, where we assume that the cellular interference at the D2D links is normalized such that noise plus interference power at each D2D receiver is equal to unity. These communication rates are achievable by using Gaussian codebooks when slot durations are large enough. }.

An application runs at the application layer of each device, and generates the bits to be stored at the transport layer queues of the devices. These bits are accepted to the network layer according to a {\em flow control} mechanism that runs at the transport layer. We let $A_i(t)$ represent the amount of data (in bits per slot) that enters the network layer at the beginning of time slot $t$ and is stored at a network layer queue with size $Q_i(t)$ at device $i$. The relationship between these important network parameters at the queue level is displayed in Fig. \ref{fig:queue_model}.  

It is assumed that the input rate $A_i(t)$ is admissible in the sense that $A_i(t) \leq A_{max}$ for all $t$, and it has a long-term average $x_i$, i.e., $x_i = \limsup_{T \rightarrow \infty} \frac{1}{T} \sum_{t=1}^T A_i(t)$. The utility obtained by the $i$th D2D pair  $U_i\paren{x_i}$ is a function of the long-term average rate $x_i$. We assume that $U_i(0) = 0$, and $U_i\paren{\cdot}$ is a continuously differentiable, monotonically increasing and concave function of its argument. This concludes the description of our system model. In the next subsections, we will formally introduce the notions of scheduling policy and network stability as well as providing some main definitions classifying/characterizing the scheduling policies and the network stability region.    

%We note that the service rate for a {\em stable} network layer queue  must be at least equal to $x_i$ \cite{Loynes62}, and that is why we %define the utility (or, satisfaction) obtained by the D2D pair $i$ from the transmission of its data as a function of $x_i$, since D2D pairs %cannot transmit more data than they already store in their queues.

%Let $A_i(t)$ represent the input rates in bits per channel use, at which data is injected in SU i in slot $t$. The rate $A_i(t)$ have long-term average $x_i$, i.e., $x_i = \lim_{T \rightarrow \infty} \frac{1}{T} \sum_{t=1}^T A_i(t)$. The utility obtained by D2D pair $i$ is a function of the long-term average rate $x_i$, since nodes cannot transmit more data than they receive. $U_i(x)$ represents the utility (or satisfaction) obtained by D2D pair $i$ from the transmission of its data, at a rate of $x$ bits per channel use.  We assume that $U_i(0) = 0$, and $U_i(\cdot)$ is continuously differentiable, monotonically increasing and concave function. The amount of traffic, $A_i(t)$ in the queues at D2D pair $i$ in slot $t$ is selected by D2D pair $i$ at the beginning of each block, and the injected information is stored in a queue with size $Q_i(t)$. The queue dynamics is illustrated in Fig. \ref{fig:queue_model}.

\subsection{Scheduling Policy}
Due to close geographical proximity of devices in our network model, only one D2D pair can communicate its data reliably over its respective wireless communication link. Hence, a scheduling decision must be made at the beginning of each time slot to select an {\em appropriate} user based on the current (both direct and interference) channel conditions.  For this purpose, roughly speaking, a scheduling policy should determine which set of links to be activated in each time slot for data transmission.  %Before giving the formal definition, let use define %$\Psi \in [0,1]^N$ is the set of all scheduling policies. 

\begin{definition}  \label{Def: Scheduling Policy} A scheduling policy $\vecbold{\cal I}: \Rp^{2N} \mapsto [0,1]^N$ is a vector-valued function $\vecbold{{\cal I}}\paren{\vecbold{h}(t),\vecbold{g}(t)} = \sqparen{{\cal I}_1\paren{\vecbold{h}(t), \vecbold{g}(t)}, \ldots, {\cal I}_N(\vecbold{h} (t),\vecbold{g}(t)) }^\intercal$ mapping the direct and interference channel states to scheduling probabilities, i.e., ${\cal I}_i\paren{\vecbold{h}(t), \vecbold{g}(t)} \in [0,1]^N$ for $i \in \brparen{1, \ldots, N}$, for devices and satisfying the feasibility constraint $\sum_{i=1}^N
{\cal I}_i\paren{\vecbold{h} (t),\vecbold{g}(t)} \leq 1$.   
\end{definition} 

It should be noted that scheduling policies given in Definition \ref{Def: Scheduling Policy} constitute a collection of {\em randomized} control mechanisms for the D2D network in question specifying scheduling probabilities for each pair of device.  Implicit in this definition, a scheduling policy does not allow two D2D pairs to be active simultaneously due to the topological constraints of our network model. More explicitly, once scheduling probabilities are identified for all D2D pairs for time slot $t$, {\em only one} of them is selected for transmission by using the probability distribution (possibly a defective one) induced by $\vecbold{{\cal I}}\paren{\vecbold{h}(t),\vecbold{g}(t)}$ over the set of device indices to determine the index of the selected D2D pair.  

%Hence, we will usually refer to the constraint $\sum_{i=1}^N
%{\cal I}_i\paren{\vecbold{h} (t),\vecbold{g}(t)} \leq 1$ in Definition \ref{Def: Scheduling Policy} as the {\em feasibility} constraint for a scheduling policy.        

An important subset of the randomized scheduling policies introduced above is the deterministic ones. We say that a scheduling policy $\vecbold{{\cal I}}\paren{\vecbold{h}(t),\vecbold{g}(t)} = \sqparen{{\cal I}_1\paren{\vecbold{h}(t), \vecbold{g}(t)}, \ldots, {\cal I}_N(\vecbold{h} (t),\vecbold{g}(t)) }^\intercal$ is a {\em deterministic} scheduling policy if ${\cal I}_i\paren{\vecbold{h}(t), \vecbold{g}(t)}$ is either zero or one for all $i \in \brparen{1, \ldots, N}$ and for all time slots $t$.  It will be shown below that the use of randomized scheduling policies will facilitate the mathematical analysis of the collection of optimization problems leading to the network stability region by turning them into convex optimization problems, {\em whilst} the solutions of these optimization problems lie in the set of deterministic scheduling policies.        

%In general, the scheduling policy given It should be noted that the above definition of scheduling policies 

%Let ${\cal I}_i\paren{\vecbold{h}(t),\vecbold{g}(t)}$ represent the scheduling decision of the $i$th pair at the joint fading state $\paren{\vecbold{h} (t),\vecbold{g}(t)}$ in time slot $t$.
%\begin{definition} We call a scheduling policy (at time slot $t$)  $\boldsymbol{{\cal I}}(\boldsymbol{h}(t),\boldsymbol{g}(t)) = \left[{\cal I}_1(\boldsymbol{h}(t),\boldsymbol{g}(t)), {\cal I}_2(\boldsymbol{h}(t),\boldsymbol{g}(t)), \ldots, {\cal I}_N(\boldsymbol{h} (t),\boldsymbol{g}(t)) \right]^\intercal$ binary if  ${\cal I}_i(\boldsymbol{h} (t),\boldsymbol{g}(t))$ is either 0 or 1 for all $i$. 
%\end{definition} 

%\begin{definition} A feasible (collision-free)
%schedule of the network is a set of D2D pairs that can transmit
%at the same time, i.e.,
%only one SU is transmitting at a given time slot, since we consider fully connected interference graph. By definition, $\sum_{i=1}^N
%{\cal I}_i((\boldsymbol{h} (t),\boldsymbol{g}(t)) \leq 1 $. 
%\end{definition}

%In randomized scheduling policy, the scheduling decision can take a number between 0 and 1 which represents the probability of selection of that schedule. Hence, it is more general compared to binary scheduling policy. In subsequent sections we define our problems considering randomized scheduling policy. However, it is shown that the optimal feasible schedule is \textit{binary}. %, which means that pairs either %"on" or "off", i.e, ${\cal I}_i(\boldsymbol{h} (t),\boldsymbol{g}(t)) \in \{0,1\}$. 

\subsection{D2D Network Stability}

In this part, we will provide the details of the notion of the D2D network stability by relating the scheduling policies to the queue level dynamics of the devices. To this end, we will first put forward the notion of interference-aware D2D operation.  All the network stability definitions presented afterwards will be with respect to the interference-aware D2D operation.    

The main communication paradigm of interest that we focus on in this paper for the coexistence of a D2D network with an AP in the same spectrum is the {\em underlay} paradigm \cite{Goldsmith09}.  The main idea underpinning the underlay communication paradigm is that the D2D network can utilize the same spectrum with the AP as long as they do not cause harmful degradation to the data communication at the AP by keeping the interference levels (peak and average) below pre-specified interference threshold values. This leads to the interference-aware D2D operation, formally defined as below.     

%In this paper, we are interested in network stability in interference-limited network. Hence, we next give the definitions of interference-limited network, network stability and stability region.

\begin{figure}
\centerline{ \includegraphics[width=3.0in]{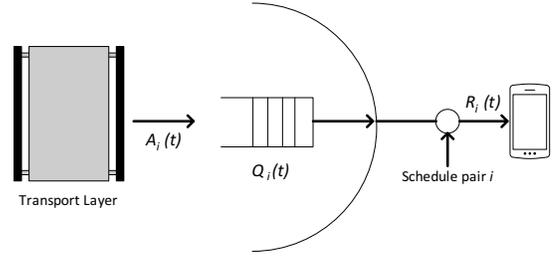}
} \caption{Queue model of pair $i$.}
\label{fig:queue_model}
\end{figure}

\begin{definition} 
A D2D network is said to be an {\em interference-aware} D2D network if the average and peak interference power levels that it causes to the AP is bounded above by the pre-specified interference threshold values as
\begin{eqnarray}
\sum_{i=1}^N\mathbb{E}_{\vecbold{h}(t),\vecbold{g}(t)}\left[P g_i(t) {\cal I}_i((\vecbold{h} (t),\vecbold{g}(t))\right] \leq \gamma
\label{eq:interference_const}
\end{eqnarray}
\vspace{-0.5cm}
\begin{eqnarray}
\sum_{i=1}^N Pg_i(t){\cal I}_i(\boldsymbol{h} (t),\boldsymbol{g}(t)) \leq \nu, 
\label{eq:interference_const_inst}
\end{eqnarray}
where $\gamma$ and $\nu$ denote the upper limits on the {\em aggregate} average and instantaneous interference powers from all D2D links, respectively. 
\end{definition}

This definition makes the coupling between the scheduling policies and the restrictions due to the interference-aware operation of the D2D network explicit.  In particular, the optimum scheduling policy achieving the maximum communication rates that can be stably supported by an interference-aware D2D network must strike a balance between choosing the best D2D link for device communication and respecting the radio etiquette rules arbitrating the spectrum access rights between devices and the AP.  The above interference constraints are primarily designed to safeguard two types of data traffic at the AP against the harmful D2D interference.  The average interference constraint in \eqref{eq:interference_const} is for delay-insensitive data traffic (e.g., text messaging) for which the messages are encoded and decoded over many slots.  On the other hand, the instantaneous interference constraint in \eqref{eq:interference_const_inst} is for delay-sensitive data traffic (e.g., video streaming) for which the messages are encoded and decoded over a single time slot. A D2D network may not know the type of data traffic at the AP at any given particular time, and hence must respect both constraints simultaneously.

Next, we formally define the concept of the stability of an interference-aware D2D network. As is standard \cite{Georgiadis}, stability here refers to being long-term averages of expected queue sizes finite, i.e., $\limsup_{T \rightarrow \infty} \frac{1}{T} \sum_{t = 0}^{T-1} \E{Q_i(t)} < \infty$. Further, we say that the D2D network is stable under $\vecbold{{\cal I}}\paren{\vecbold{h}(t),\vecbold{g}(t)}$ if the network layer queues of all devices are stable. An important concept that expands upon definition of network stability and relates the flow control and scheduling mechanisms for a D2D network is the network stability region, which is defined as below.  
%\begin{definition} \label{Def: Network Stability}
%Consider an interference-aware D2D network operating according to a feasible scheduling policy $\vecbold{{\cal I}}\paren{\vecbold{h}(t),\vecbold{g}(t)}$ that satisfies the interference constraints in \eqref{eq:interference_const} and \eqref{eq:interference_const_inst}.  Then, we say the network layer queue $Q_i(t)$ at device pair $i$ is stable under  $\vecbold{{\cal I}}\paren{\vecbold{h}(t),\vecbold{g}(t)}$ if
%\begin{eqnarray}
%\limsup_{T \rightarrow \infty} \frac{1}{T} \sum_{t = 0}^{T-1} \E{Q_i(t)} < \infty. 
%\end{eqnarray}
%Further, we say that the D2D network is stable under $\vecbold{{\cal I}}\paren{\vecbold{h}(t),\vecbold{g}(t)}$ if the network layer queues of all devices are stable. 
%\end{definition}

%Throughout this paper, we restrict our attention to the strong stability definition given above, and often use the term stability to refer to strong stability.

%It is known(e.g., [26]) that the capacity region is given by
% = { |   0 and 9μ 2 Co(M), < μ}, (2)
%where Co(M) is the convex hull of the set of feasible
%schedules in M. 

%We use the indicator variable ${\cal I}_i(t)$ to represent
%the scheduler decision:
%\begin{equation}
%{\cal I}_i(t)=\begin{cases} 1, & \text{information transmitted from D2D pair $i$} \\
%0, & \text{otherwise}
%\end{cases}.
%\end{equation}

\begin{definition} \label{Def: Network Stability Region}
The {\em network stability region} of an interference-aware D2D network, denoted by $\Lambda$, consists of all arrival rate vectors $\vecbold{x} = \paren{x_1, x_2, \ldots , x_N}^\intercal$ such that there exists a scheduling policy $\boldsymbol{{\cal I}}\paren{\boldsymbol{h}(t),\boldsymbol{g}(t)}$ satisfying the conditions below for all $i \in \brparen{1, \ldots, N}$.

\vspace{-0.15in}
%\small
\begin{align}
&\mathbb{E}_{\vecbold{h}(t),\vecbold{g}(t)}\left[{\cal I}_i(\boldsymbol{h} (t),\boldsymbol{g}(t)) R_i(t)\right] \geq x_i, \label{eq:stab_const} \\
&\sum_{i=1}^N\mathbb{E}_{\vecbold{h}(t),\vecbold{g}(t)}\left[P g_i(t){\cal I}_i(\vecbold{h}(t),\vecbold{g}(t))\right] \leq \gamma, \label{eq:int_const} \\
&\sum_{i=1}^N P g_i(t){\cal I}_i((\vecbold{h} (t),\vecbold{g}(t)) \leq \nu, \mbox{ and }\sum_{i=1}^N {\cal I}_i((\vecbold{h} (t),\vecbold{g}(t)) \leq 1.   \label{eq:sched_const}
\end{align}
\end{definition}		

\normalsize
The constraints in \eqref{eq:int_const} and \eqref{eq:sched_const}, they are due to the interference-aware operation of the D2D network, and the feasibility condition.  The constraint in \eqref{eq:stab_const} is the classical necessity constraint for the queue stability describing the fact that the incoming rate to the network layer queues should be equal to or smaller than the outgoing service rate, which depends on the choice of the scheduling policy in our particular D2D communication scenario \cite{Neely05}.\footnote{Note that $\Lambda$ is the minimal set that contains all achievable arrival rates, i.e., no vector $\vecbold{x}$ outside $\Lambda$ can be stabilized by any feasible and interference-aware scheduling policy. Although not needed in our analysis, it is also important to note that  it can be easily shown that the stability region $\Lambda$ is a convex set by using the standard time-averaging arguments.}

In the next section, we will obtain the Pareto boundary of the network stability region, where no feasible and interference-aware scheduling policy can stabilize the D2D network when the arrival rates are beyond this boundary. This will provide a complete characterization of $\Lambda$. This characterization will be carried out under the full channel-state information (CSI) assumption.  Although helpful in understanding the maximum rates that can be stably supported by a D2D network, such a characterization of the D2D network stability region does not provide us with any insights regarding how to design dynamic control mechanisms achieving the rates in $\Lambda$.  

To resolve this drawback, we design a dynamic but centralized flow control and scheduling algorithm that achieves all the rates within the D2D network stability region in Section \ref{control}. The scheduling part of the proposed algorithm provides design insights into how to construct a feasible and interference-aware scheduling mechanism for a D2D network. In addition to stabilizing an interference-aware D2D network, the proposed algorithm also maximizes the collective utility of the devices. The flow control part of the proposed algorithm provides design insights into how to construct flow control mechanism to maximize collective network performance. The distributed solutions achieving these desirable properties up to some provable performance bounds are given in Section \ref{sec:dist_scheduling_selective}.     

%When we analyze the stability region, we assume that every user has full CSI of its direct and interference channels, i.e., $h_i(t)$ and $g_i(t)$ are available to D2D pair $i$ at every block $t$. Then, we design a dynamic algorithm to achieve the rates within the stability region. Indeed, the proposed dynamic algorithm performs joint flow control and scheduling which not only stabilizes the queues of D2D network, but also maximizes the collective utility of D2D network.  We give centralized and distributed solutions to the problem, and derive performance bounds of the algorithms.

 %Thus, the achievable individual and sum rates we
%derive constitute upper bounds on the achievable rates with partial
%CSI, subject to the perfect privacy constraint.
%On the other hand, when we formulate our problem as that of network
%utility maximization problem, we only assume knowledge of
%instantaneous channel gains without requiring the knowledge of prior
%distribution of channel gains. %Hence, private encoding is performed
%over a single block length unlike the case when achievable rates are
%calculated. Clearly, privacy rate attained with this scheme is lower
%than the achievable rate obtained with full CSI.
}
{\allowdisplaybreaks\section{Stability Region for Interference-Aware D2D Network}
\label{sec:ach_int_limited}
%Consider two users scenario in which two SUs are transmitting
%information over their uplink channels resulting interference to the
%CBS. Here, we consider two different network types as interference
%limited and power-interference limited networks. In interference
%limited network, SUs' transmissions are limited by an average
%interference constraint, i.e., we required that expected value of
%interference to the AP due to transmissions of SUs is below a given
%threshold, $\gamma$. In power-interference limited networks, SUs'
%transmissions are limited by both individual power constraints of
%SUs and an average interference constraint that they cause to AP.
%In both network types, we intend to study the effect of interference
%constraint on the achievable rates.

In this section, we derive the boundary of the stability region of an
interference-aware D2D network such that any arrival rate vector, $\vecbold{x}$ outside the closure of the boundary is unattainable. Then, we analyze the effect of interference-aware communication on the network stability region by comparing the boundaries obtained with and without interference constraints.

 %for the cases when the scheduling is
%performed by a centralized entity (centralized algorithm) %or when
%the nodes decide whether to transmit or not based on their local
%knowledge (distributed algorithm). 
%In an interference limited network, D2D transmissions are limited by an average interference
%constraint at the AP, i.e., we require that the long-term average of interference to
%the AP due to transmissions of D2D pairs is kept below a given threshold,
%$\gamma$. 

We begin our analysis by computing the boundary of network stability region. This is equivalent to maximizing the average outgoing (service) rate achieved by $i$ for given average service rate values of other devices. Recall that the average arrival rate should be smaller or equal to the average service rate in a stable network. Hence, we say that any arrival rate of device $i$, $x_i$, that is larger than this maximized service rate, cannot be achieved. Before giving the mathematical description of the problem, we make following remark. Since we assume that the channels are ergodic and stationary, we utilize the statistical averages in constructing the optimization problem. Hence, we ignore the time index for the sake of notational simplicity in this section. But in the next section, when we perform dynamic control, we again introduce time index back. Further, for notational convenience, let $\Psi \in [0,1]^N$ be the set of all scheduling policies. Therefore, the aim is to maximize $\E{{\cal I}_i(\boldsymbol{h},\boldsymbol{g})R_i}$, associated with the point $\E{{\cal I}_j(\boldsymbol{h},\boldsymbol{g})R_j}
= \alpha_j, \forall j \neq i$, by solving the following
linear program:

%Recall that ${\cal I}_i(\boldsymbol{h}(t),\boldsymbol{g}(t))$ is the scheduling decision for pair $i$ in slot $t$. 
%\subsection{Achievable Rate of Interference Limited Network}
 %The reason of using constant transmit
%power is to be able to make a fair comparison with the case without
%interference constraint, since the achievable rates with varying
%power are unbounded without power and interference constraint.
%Recall that ${\cal I}_i(t)$ as the indicator variable, which takes
%on a value 1, if D2D pair $i$ transmits its information over block $t$ and
%0 otherwise. Then, the average rate achieved by D2D pair $i$ is $R_i^\text{avg} =
%\E{{\cal I}_i(t)R_i(t)}$.  

% To find the pareto boundary of the stability region, we maximize the average rate achieved by a pair by fixing the average rates achieved by %other users. That is to say,
\vspace{-0.15in}
\begin{align}
    \max_{\boldsymbol{{\cal I}} \in \Psi} &\E{{\cal I}_i(\boldsymbol{h},\boldsymbol{g})R_i}\label{eq:obj_linear}\\
    \mbox{subject to } & \E{{\cal I}_j(\boldsymbol{h},\boldsymbol{g})R_j} \geq \alpha_j, \ \forall j \neq i \label{eq:obj_trans} \\
    & \E{\sum_{l=1}^N P{\cal I}_l(\boldsymbol{h},\boldsymbol{g}) g_l} \leq
    \gamma, \label{eq:obj_int} \\
		&\sum_{l=1}^N Pg_l{\cal I}_l(\boldsymbol{h},\boldsymbol{g}) \leq \nu \mbox{ and } 		\sum_{l=1}^N
{\cal I}_l(\boldsymbol{h},\boldsymbol{g}) \leq 1,
\label{eq:sch_const}
\end{align}
where the expectations are over the joint distribution of the
instantaneous channel gains of direct and interference channels. %Note that, here we consider that the scheduling decision can take a value %between 0 and 1, . Then, the scheduling decision corresponds to the probability of given the particular %scheduling decision. 
We solve the above optimization problem using the dual method that is particularly
appealing to our problem structure, whose solution is given in the next theorem. %Since our problem is a convex problem, it can be shown that %the optimal solution of the dual problem is exactly the same as the solution of the original problem, i.e., the duaility gap is zero %\cite{Boyd}.

%programming problem has a dual form. The original problem is sometimes called
%the primal problem. It can be shown that the optimal solution of the dual of any convex problem is
%exactly the same as the solution of the primal problem

\begin{theorem}
The solution of the optimization problem given in \eqref{eq:obj_linear}-\eqref{eq:sch_const} is given by
\begin{align}
{\cal I}_j^*(\boldsymbol{h},\boldsymbol{g})=\begin{cases} 1, & \text{ if  }W_j = \max_{1\leq l \leq N} W_l, \ \forall l \neq j \text{ and } W_j \geq 0 \\  &\text{     and } Pg_j \leq \nu \\
0, & \text{otherwise}
\end{cases}
\end{align}
where $W_j =  \lambda_j^*R_j - \mu^*Pg_j$ for all $j \neq i$ and $W_i =  R_i - \mu^*Pg_i$, and
$\lambda_j^*$ and $\mu^*$ are Langrange multipliers associated with the rate and interference constraints in \eqref{eq:obj_trans} and \eqref{eq:obj_int} for which $\E{{\cal
I}_j^*(\boldsymbol{h},\boldsymbol{g}) R_j} = \alpha_j$, for all $j \neq i$ and the constraint in \eqref{eq:obj_int} is satisfied.

%is the value of $\lambda_j$ for which $\E{{\cal
%I}_j^*(\boldsymbol{h},\boldsymbol{g}) R_j} = \alpha_j$. %, since $\lambda^*\left(\E{{\cal I}_j(\boldsymbol{h},\boldsymbol{g})R_j} -
%\alpha_j\right)\geq 0$. 
%Similarly the value of $\mu^*$ is the value
%of $\mu$ for which $\E{P \sum_{l=1}^N {\cal I}_j(\boldsymbol{h},\boldsymbol{g})g_j} = \gamma$.

\label{thm:optimalscheduling}
\end{theorem}

\begin{IEEEproof} Please see Appendix \ref{proof:optimalscheduling}. \end{IEEEproof} 

%To be more formal, we can use functional optimization. Readers are referred to texts \cite{Luo05} for comprehensive results on functional optimization.

Theorem \ref{thm:optimalscheduling} gives us the optimal scheduling policy $\vecbold{{\cal I}}^*$ achieving $\E{{\cal
I}_j^*(\boldsymbol{h},\boldsymbol{g}) R_j} = \alpha_j$ for all $j\neq i$. Then, the boundary of the stability region can be attained by varying $\alpha_j, \forall j\neq i$, and obtaining the points where the average rates of device $i$ are maximized. Another important point is that even though we state the optimization for the randomized scheduling policies, i.e., ${\cal I}_j(\boldsymbol{h},\boldsymbol{g})  \in [0,1 ]$, the optimal solution turns out to be a deterministic scheduling policy, i.e., ${\cal I}_j(\boldsymbol{h},\boldsymbol{g}) $ is either zero or one. In addition, observe that if the condition $P g_j > \nu$ and $W_j < 0$ for all $i$, then the channel remains idle, i.e., $\sum_{l=1}^N {\cal I}_l(\boldsymbol{h},\boldsymbol{g})  = 0$. The reason is that the channel conditions are not good
enough to access the channel at the expense of the interference caused to the transmission of AP.

%Observe that ${\cal
%I}_l(\boldsymbol{h},\boldsymbol{g}) = 1, {\cal I}_j(\boldsymbol{h},\boldsymbol{g}) =
%0, \ \forall j \neq l$ if the objective function on \eqref{eq:obj_fun_sep} is maximized for the
%transmission of $l$th D2D pair, or it will choose ${\cal
%I}_l(\boldsymbol{h},\boldsymbol{g}) = 0, \ \forall l$ otherwise. 

%The solution of above problem is on the boundary corresponding to binary scheduling policy. That is to say, $The last
%condition suggests that 

As indicated above, there may be time instants during which the channel remains idle in an interference-aware D2D network to safeguard the AP. This will result in a decrease in optimal rates due to under-utilization of the channel. Consequently, it leads to the contraction of the network stability region. To understand this phenomenon better, we also derive the optimum scheduling policy without interference
constraints, and compare the achievable rate regions of both cases with and
without interference constraints. Following similar arguments above, we have the following optimum scheduling problem

\vspace{-0.15in}
\begin{align}
    \max_{\boldsymbol{{\cal I}} \in \Psi} &\E{{\cal I}_i(\boldsymbol{h},\boldsymbol{g})R_i} \label{eq:obj_wo_int}\\
    \mbox{subject to } & \E{{\cal I}_j(\boldsymbol{h},\boldsymbol{g})R_j} \geq \alpha_j, \ \forall j \neq i  \mbox{ and } \sum_{l=1}^N
{\cal I}_l \leq 1
\label{eq:obj_trans1}		
\end{align}
without interference constraints, whose solution is given by the next lemma.

\begin{lemma}
The solution of the optimum scheduling problem in \eqref{eq:obj_wo_int}-\eqref{eq:obj_trans1} is given as:
\begin{align}
{\cal I}_j^*(\boldsymbol{h},\boldsymbol{g})=\begin{cases} 1, & \text{ if  } \lambda_j^*R_j > \lambda_l^*R_l   \ \forall l \neq j  \\
0, & \text{otherwise}
\end{cases}
\end{align}
\end{lemma}
where $\lambda_j^*$ is the Lagrange multiplier associated with the rate constraint in \eqref{eq:obj_trans1}.

\begin{IEEEproof}   The proof follows the similar lines with the proof of Theorem \ref{thm:optimalscheduling}, and hence is skipped to avoid repetitions. \end{IEEEproof} 
%For Gaussian channels described in Section~\ref{sec:model}, we
%obtain:

%\begin{align}
%{\cal I}_j=\begin{cases} 1, & \text{ if  } (1+Ph_{j})^{\lambda_j^*}  > (1+Ph_l)^{\lambda_l^*}  \ \forall l \neq j  \\
%0, & \text{otherwise}
%\end{cases}
%\end{align}

%{\bf Remark:} The same problem can be written by assuming that the scheduling decision only takes value 0 or 1. Then, the problem becomes% %integer problem and can be solved by using Lagrangian relaxation. However, the problem should satisfy several conditions (can be found in \cite{}) to guarantee that the solution of the original problem% is the same with the solution of the Langrange problem.
%\begin{figure}
%\includegraphics[width=3.5in]{dec_region.pdf}
%\label{fig:dec_region}   \caption{The decision regions for two users}
%\end{figure}

%The associated solution ${\cal I}^*$ is graphically illustrated on the
%$(h_1,h_2)$ space in Fig. \ref{fig:dec_region} for P = 1 and $g_1=g_2=g$. When the value of $\lambda = 1$, the optimal decision region for
%${\cal I}_2 = 1$  is the upper half of the first quadrant
%represented by $h_2 > h1$. As the value of $\lambda$ increases the optimal decision region for ${\cal I}_2 = 1$ starts to span the entire first quadrant, i.e., all
%$h_1,h_2 > 0$. Furthermore, the shaded area is the difference between the optimal decision between the problems with and without interference constraint. In interference-constrainted problem, no user %transmits in shaded area. If the interference constraint is inactive, i.e., $\mu^* = 0$, the solution regions of both problems converges to the same region. 

\textit{Interpretation of Stability Network Region for Two Device
Case: }
\begin{figure*}
\centerline{ \subfloat[Network stability region for varying interference
parameter,
$\gamma$]{\includegraphics[width=3.0in]{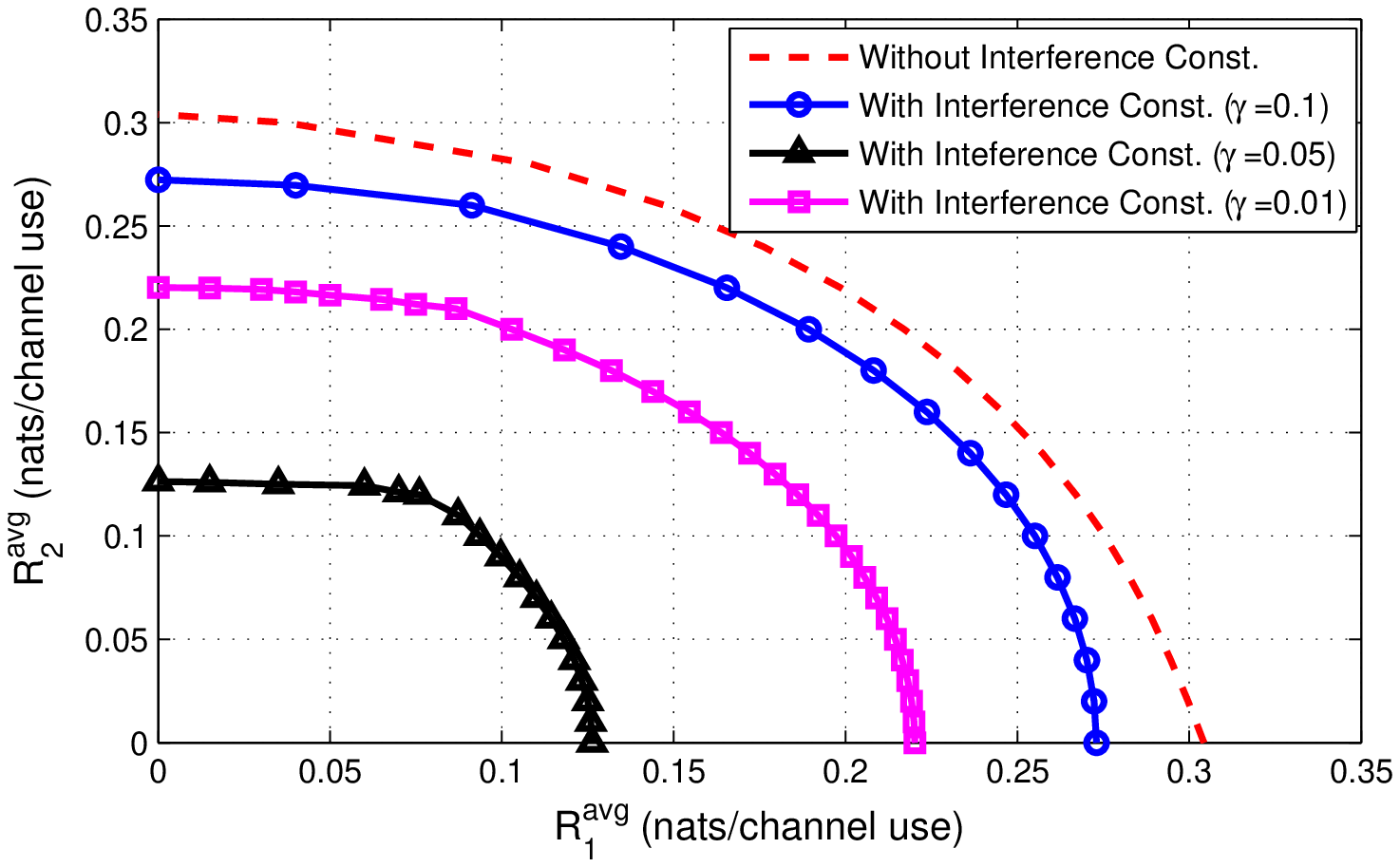}
\label{fig:gamma}} \subfloat[Network stability region for varying
interference channel gain,
$\E{g_2}$]{\includegraphics[width=3.0in]{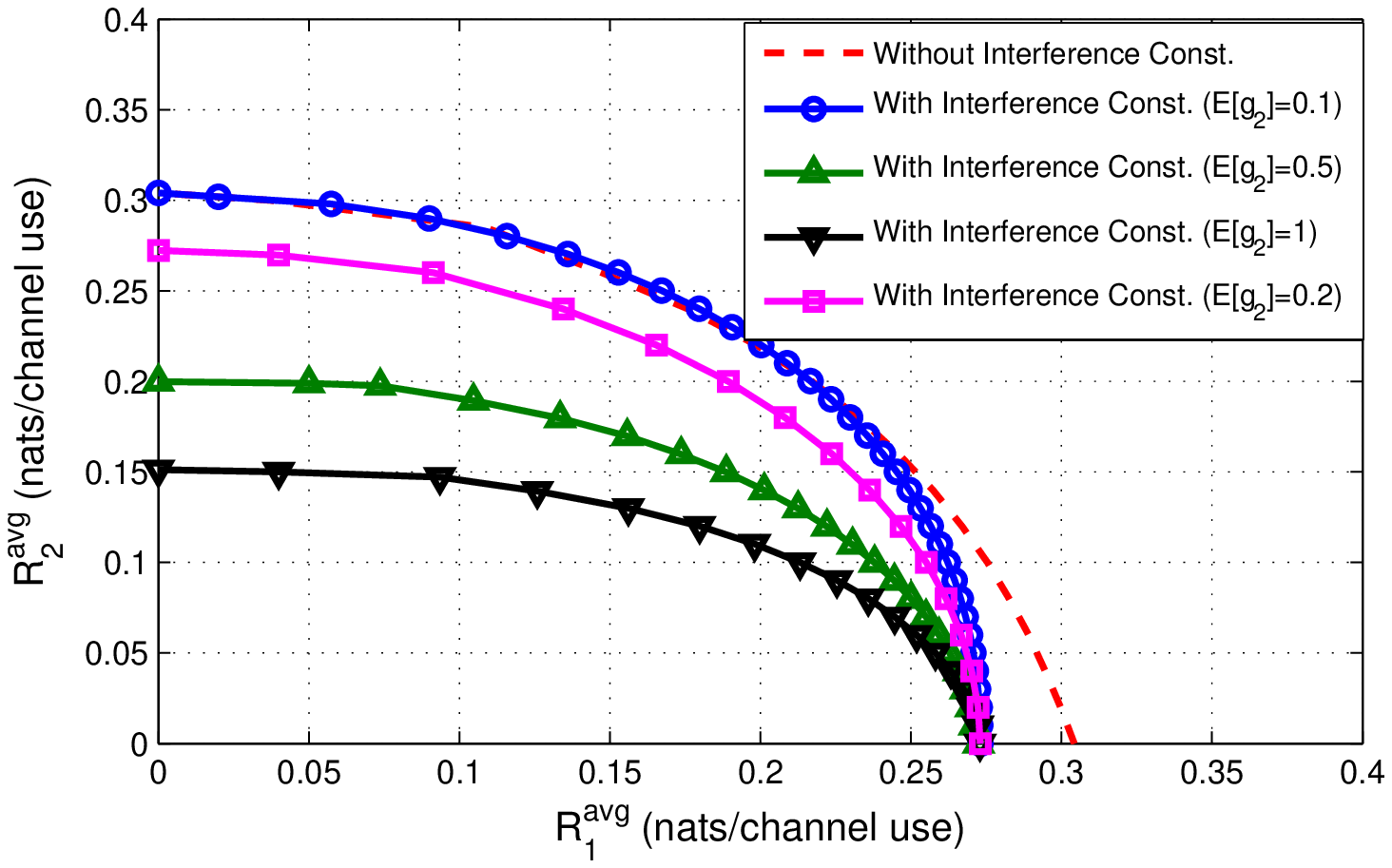}%
\label{fig:int_ch}  }} \caption{Network stability region for the communication scenario with two D2D pairs, where $R_i^{avg} = \E{{\cal I}_i^*R_i}$. }
\end{figure*}
Now, we consider a communication scenario containing only two D2D pairs. In this case, the interference-aware optimum scheduling policy is given by 

\vspace{-0.1in}
\small
\begin{align*}
{\cal I}_1(\boldsymbol{h},\boldsymbol{g})=\begin{cases} 1, & \text{ if  } R_1 - \mu^*Pg_1 > \lambda^* R_2 - \mu^*Pg_2 \text{ and } R_1 > \mu^*Pg_1 \\
0, & \text{otherwise}
\end{cases}
\end{align*}
\normalsize
and
\small
\begin{align*}
{\cal I}_2(\boldsymbol{h},\boldsymbol{g})=\begin{cases} 1, & \text{ if  } R_1 - \mu^*Pg_1 < \lambda R_2 - \mu^*Pg_2 \text{ and } \lambda^* R_2 > \mu^*Pg_1 \\
0, & \text{otherwise}
\end{cases},
\end{align*}

\normalsize

In Figs.~\ref{fig:gamma}-\ref{fig:int_ch}, the stability region for a two-link D2D network, is illustrated for Rayleigh fading direct and interference channels. To plot the regions, we varied the rate achieved by the second D2D pair, and calculated $\lambda^*$ and
the boundary rate pair for each point. Recall that $h_i$ and $g_i$ are the direct and interference channel gains of $i$th D2D pair, respectively. In Fig. \ref{fig:gamma}, we take
$\E{h_1}=\E{h_2} = 0.4$ and $\E{g_1}=\E{g_2} = 0.2$ and we obtain
different boundary rate pair for varying interference parameter,
$\gamma$. %In simulations, we observe that the
%rate regions with and without interference constraint coincide for $\gamma = 0.25$. The
%reason is that the interference constraint in
%\eqref{eq:obj_int} becomes inactive in this case, i.e., $\mu^* = 0$. Thus, the
%optimal solutions for both problems are the same. 
As seen in Fig. \ref{fig:gamma}, as we decrease
$\gamma$ value, i.e., the interference constraint is more stringent, both
D2D pairs have smaller transmission opportunities to meet the interference
constraint. Thus, the network stability region becomes smaller. In Fig. \ref{fig:int_ch}, we fixed the value of $\gamma$
at 0.1 and vary $\E{g_2}$. As seen in Fig. \ref{fig:int_ch}, when $\E{g_2} =
0.1$,  the network stability regions (with and without interference
constraint) coincides for smaller values of $R_1^{avg}$, where the second D2D pair 
takes a higher portion of transmissions.  In this case, the interference constraint is inactive since the second D2D pair with smaller interference channel gain transmits predominantly. On the other hand, as $\E{g_2}$ increases, i.e., $\E{g_2}=0.2, 0.5, 1 $, more
interference is caused to the AP and the network stability region shrinks.

}
{\allowdisplaybreaks\section{Control of Underlay D2D Communication Networks with Centralized Scheduling}
\label{control}

In the previous section, we characterize the stability region by obtaining maximum rates that an interference-aware D2D network can support. In this section, we will present a dynamic control algorithm  that will solve a network utility maximization (NUM) problem while stabilizing the network layer queues in a D2D network. To do so, we follow a cross-layer design approach. In the lower layer, the scheduling policy ensures network stability and satisfies the interference requirements. In the upper layer, on the other hand, flow control policy strives to move the network layer arrival rates to the optimal point within the stability region. Since the derived cross-layer algorithm will be a dynamic online algorithm, we will use the time index $t$ in this section again to indicate its operation in time.
%that opportunistically schedules the D2D pairs with the objective of
%maximizing total expected utility of the network subject to
%interference constraint to the AP %and individual power constraint
%of D2d pairs 
%while maintaining the stability of queues of D2D pairs.
%In the previous sections, the achievable rate region of the network
%is calculated based on the assumption that complete distributions of
%uplink and interference channel gains are available. In this
%section, we assume that the nodes are capable of obtaining perfect
%instantaneous CSI, but do not have access to the prior distributions
%of the channel states. 

The
dynamic cross-layer algorithm takes the queue lengths (
both virtual and real queues) and instantaneous direct- and
cross-channel gains as input, and determines the scheduled device $i$ at each time slot as output.
We start our analysis by first formulating the NUM problem and providing the queue dynamics to set the stage for the cross-layer design approach. 

\subsection{Problem Formulation}

Our objective is to stabilize the network while maximizing the
sum of device utilities. That is, we aim to find the solution
of the following NUM problem:

\vspace{-0.15in}
\begin{align}
    \max \ &\sum_i U_i(x_i) \label{eq:opt-objective} \\
    \mbox{ subject to } & x_i \in \Lambda \label{eq:const-stability} %\\
    %& \E{P_i(t)} \leq \beta_i \ \forall i \label{eq:const-power}\\
    %& \sum_{i=1}^N \E{Pg_i(t)} \leq \gamma \label{eq:const-interference}
\end{align}

The objective function in \eqref{eq:opt-objective} calculates the
total expected utility of D2D pairs over random stationary channel
conditions and scheduling decisions. The constraint
\eqref{eq:const-stability} ensures that network layer arrival rates of D2D pairs are
within the rate region supported by the network defined as
$\Lambda$. The above problem could in principle be solved by means of standard convex optimization techniques if the stability region is known in advance. Although this approach may give us an idea about how to select transmission rates, it will not say anything about how we can reach the optimum operating by relating the solution to the design of wireless networks. Thus, in the following subsections, we develop a practical dynamic control algorithm to facilitate our understanding of the interplay between interference requirements of the D2D network and the critical functionalities of wireless networks, such as scheduling, and flow control.

\subsection{Queue Dynamics}

We assume that there is an infinite backlog of data at the transport
layer of each node. Our proposed dynamic flow control algorithm
determines the amount of traffic injected into the queues at the
network layer. The dynamics of the network layer queue of $i$th D2D pair is given as
follows:

\vspace{-0.15in}
\begin{align}
Q_i(t+1)=\left[ Q_i(t)-{\cal I}_i(t) R_i(t)\right]^+ + A_i(t).
\label{eq:real_queue}
\end{align}

To meet the average interference constraint given in \eqref{eq:interference_const}, we also maintain
a virtual queue as:

\vspace{-0.15in}
\begin{align}
Z(t+1) &= \left[ Z(t)- \gamma  + \sum_{i=1}^N {\cal I}_i(t)\cdot
Pg_i(t)\right]^+ \label{eq:interference_queue}
\end{align}

The state
of the virtual queue at any given time instant is an indicator on
the amount by which we exceed the allowable interference
constraint. Thus, the larger the state of these queues, the more
conservative our dynamic algorithm has to get towards meeting
these constraints, i.e., the less transmissions will take place by D2D pairs. The strong stability of virtual queues guarantees that the interference constraints are satisfied in the long run (see Theorem 5.1 in \cite{Georgiadis}).

%that yields a resulting matrix of throughput r is arbitrarily close to the optimal solution of ()-()

\subsection{Dynamic Control}

The proposed cross-layer dynamic control algorithm is based on the
stochastic network optimization framework \cite{Georgiadis}. This
framework allows the solution of a long-term stochastic optimization
problem without requiring explicit characterization of the
stability region. Furthermore, it enables the simultaneous treatment of stability and performance optimization  by the introduction of virtual queues to transform performance constraints into queuing stability problems. 

To this end, consider queue backlog vectors for D2D communication pairs as
$\Qv(t)=(Q_1(t),\ldots, Q_N(t))$
and $Z(t)$. Let $L(\Qv(t),Z(t))$ be a quadratic Lyapunov
function of real and virtual queue backlogs defined as:

\vspace{-0.15in}
\begin{equation} L(\Qv(t),Z(t)) =
\frac{1}{2} \left((Z(t))^2 +
\sum_{i=1}^N(Q_i(t))^2 \right).
\label{eq:lyapunov-function}
\end{equation}

Also, consider the one-step expected Lyapunov drift, $\Delta(t)$ for
the Lyapunov function as:
\begin{equation*} \Delta(t) =
\mathbb{E}\left[ L(\Qv(t+1),Z(t+1)) -
L(\Qv(t),Z(t))|\Qv(t),Z(t)\right].
%\label{eq:lyapunov-drift}
\end{equation*}

%Note that $\Delta(t) = 0$ if and only if all network queues are empty at
%time $t$, and that $\Delta(t)$ is large whenever one or more components in $\Delta(t)$ is large. 
The aim of stochastic optimization framework is to minimize the drift to ensure the network stability, which can be achieved by having negative Lyapunov drift
whenever the sum of queue backlogs is sufficiently large. Intuitively, this
property ensures network stability because whenever the queue backlog
vector leaves  the stability region, the negative drift eventually drives it back to this region. Furthermore, the following  utility-mixed Lyapunov drift
\vspace{-0.1in}
\begin{equation} \Delta^U(t)=\Delta(t) - V\E{\sum_{i=1}^N
U_i(A_i(t)) | \Qv(t),Z(t)}, \label{eq:deltawithreward}
\end{equation}
enables us to maximize the network performance in conjunction with the network stability, where the conditional expectation is taken with respect to the random
fading realizations. 

Next, we present the control algorithm that minimizes \eqref{eq:deltawithreward} and provide its optimality in Theorem \ref{thm:optimalcontrol}.

\noindent {\bf Control Algorithm:} Making an analogy to back pressure algorithm, we propose the following cross-layer algorithm which executes the following steps in each time $t$:
\begin{enumerate}
\item[\bf (1)] {\bf Upper Layer - Flow control:} The flow controller at each device observes its current queue backlog, $Q_i(t)$. It then injects $A_i(t)$ bits, where $A_i(t)$ is the solution of the following optimization:
\begin{align}
A_i(t)=\argmax_{ 0 \leq x \leq A_{max} }\{ V U_i(x)- Q_i(t)x \},
\end{align}
where $V > 0$ is a design constant that  will determine the final performance of the designed algorithm. The above identity involves maximizing a concave function, which can be easily solved
by using  convex optimization techniques \cite{Boyd}.

\item[\bf (2)] {\bf Lower Layer - Scheduling:} A scheduler observes the backlogs in all devices, and all fading states. Then, it determines the scheduling decision for time slot $t$, $\boldsymbol{{\cal I}}(t)$ as the solution of the following optimization:

%D2D pair $i$ among the pairs that satisfies the instantaneous interference constraint, i.e., $Pg_i(t) \leq \nu$, and satisfy where
%\[ (\boldsymbol{{\cal I}}(t))=\argmax_{\boldsymbol{{\cal I}} \in \Psi }\ \left\{ \sum_i {\cal I}_i W_i(t)\right\}, \]
\vspace{-0.15in}
\begin{align*}
\boldsymbol{{\cal I}}(t) = &\argmax_{\boldsymbol{{\cal I}} \in \Psi }\ \left\{ \sum_{i=1}^N {\cal I}_i W_i(t)\right\} \\
\mbox{subject to } 	&\sum_{i=1}^N {\cal I}_i W_i(t) \leq 1 \mbox{ and } {\cal I}_i Pg_i(t) \leq \nu, \ \forall i,	
\end{align*}
%\vspace{-0.1in}
where $W_i(t)$ is the weight of pair $i$ and is given as: 
\begin{align}
	W_i(t)= Q_i(t){\cal I}_i(t)R_i(t)-Z(t)Pg_i(t).
	\label{eq:weight}
\end{align}
The above is a standard linear optimization problem, whose solution is obtained on the boundary. Specifically, among the pairs that satisfies the instantaneous interference constraint, and has the maximum weight, is allowed to transmit at a given time slot.
\end{enumerate}

%we again show that the optimal scheduling policy is deterministic meaning that only one device is allowed to transmit at given time slot
 
We note that the parameter $V$ in the flow control algorithm that determines the extent to which the utility optimization is emphasized. Indeed,
if $V$ is large relative to the current backlog in the source queues, then
the admitted rates $A_i(t)$ will be large, increasing the time average
utility while consequently increasing congestion. This effect is mitigated as the backlog grows at the source queues and flow control decisions become more conservative. Note that the flow control algorithm is decentralized because the control values for each node require only knowledge of the queue backlogs in device $i$. 

In the scheduling policy, the weight equation in \eqref{eq:weight} consists of reward $Q_i(t) R_i(t)$ and cost $P Z(t) g_i(t)$ terms. Specifically, the
larger the data queue backlog size $Q_i(t)$ and/or higher the instantaneous channel rate $R_i(t)$, the more likely the transmission of D2D pair
$i$ occurs. On the other hand, larger the interference queue
backlog $Z(t)$ (representing the interference level caused to AP)
and/or higher the interference channel gain $g_i(t)$, the less
likely the transmission of pair $i$ takes place. In this setting, the flow control algorithm strives to maximize collective network utility, whereas the scheduling policy makes sure that the utility maximizing operating point is within the stability region. Indeed, by utilizing the proposed scheduling algorithm, we can achieve any point in the stability region.

\begin{theorem}
\label{thm:optimalcontrol}
Suppose $\boldsymbol{x^*} = [x_1^*, \ldots, x_N^*]$ is the average arrival rates produced by the proposed dynamic control algorithm. Then, for any flow parameter  $V > 0$,  the dynamic control algorithm yields the following performance bound for the aggregate network utility:
 \begin{align*}
    \sum_{i=1}^N  U_i(x_i^*) &\geq U^* - \frac{B_1}{V}
		\end{align*}		
while bounding the total long-term expected queue lengths as:	
\begin{align*}
    \limsup_{T \rightarrow \infty} \frac{1}{T} \sum_{\tau =
    0}^{T-1}\sum_{i=1}^N \E{Q_i(\tau)} &\leq
    \frac{B_1+V\kappa}{\epsilon_1},
 \end{align*}
 where $B_1,\epsilon_1,\kappa > 0$ are constants, and $U^*$ is the optimal
 aggregate utility, i.e., the solution of the problem in
 \eqref{eq:opt-objective}-\eqref{eq:const-stability}. This theorem shows that the proposed dynamic control gets arbitrarily close to the
optimal utility by choosing $V$ sufficiently large at the expense of
proportionally increased average queue sizes.
\end{theorem}

 \begin{IEEEproof}  The proof is given in Appendix \ref{proof:optimalcontrol}. \end{IEEEproof}

%According to Theorem \ref{thm:optimalcontrol}, there is a trade-off in choosing the
%parameter V, i. e., larger values achieve a solution closer to
%the optimal, but at the same time increases the aggregate queue
%length. 

We note that the proposed dynamic control algorithm is not distributed since its scheduling part depends on global queue length information. As
compared to the distributed scheduling algorithms, the centralized
scheduling schemes usually lead to a better performance at the cost
of requiring a central authority to allocate the network resources.
In a large-scale wireless network, such a central authority does not
always exist. Furthermore, implementation of the centralized
algorithms results in high overhead on the network due to collecting
channel conditions and queue states of all users. In the remainder of the paper, we focus on designing distributed algorithms relaxing the
assumptions necessary for the centralized algorithm. Note that the flow
control part of the proposed solution is already distributed, i.e, each node decides its admitted flow
by only local information. Thus, they remain the same below.
}
{\allowdisplaybreaks\section{Channel-Aware Distributed Algorithms }
\label{sec:dist_scheduling_selective}

In this section, we relax the requirement of having a central authority for device scheduling in Section \ref{control} by investigating contention based  distributed scheduling algorithm with multiple round contention, called Channel-Aware Distributed Schedulers (CADS) with uniform mapping, that operates based on the local queue size and channel state information at each D2D pair. The distributed mode of operation necessitates the modification of the NUM problem as

%For that reason, a distributed scheduling
%algorithm would be very appealing even if it could only obtain a
%certain fraction of the achievable rate region of a centralized
%scheduling scheme. %Hence, in this section, we aim to design practical distributed algorithms with good performance guarantees.

%Due to limited information about the state of the network, obtaining the maximum weight at each time slot may not be feasible due to limited information, resulting in imperfect (sub-optimal) solution. This leads to reduction in stability region, i.e., distributed scheduling can only stabilize the rates within some portion of stability region. That is why, we modify the optimization problem for distributed scheduling as:
\vspace{-0.15in}
\begin{align}
    \max \ &\sum_i U_i(x_i) \label{eq:opt-objective_dist} \\
    \mbox{ subject to } & x_i \in \alpha \Lambda \label{eq:const-stability_dist} %\\
    %& \E{P_i(t)} \leq \beta_i \ \forall i \label{eq:const-power}\\
    %& \sum_{i=1}^N \E{Pg_i(t)} \leq \gamma \label{eq:const-interference}
\end{align}
where $\alpha$ is the contraction coefficient. The constraint in \eqref{eq:const-stability_dist} suggests that
the distributed scheduling algorithms can still stabilize the network, provided that the arrival rates are interior
to $\alpha\Lambda$, which is a $\alpha$ scaled version of the stability region.

In the remainder of the section, we will introduce a distributed scheduling algorithm that is channel-aware in the sense that they can utilize the diversity gain in time-varying environments, and obtain its performance bounds by characterizing $\alpha$.  For analytical purposes, we assume that all channel gains, $h_i(t)$ and $g_i(t)$ are i.i.d. However, we perform simulations for i.i.d and non-i.i.d. cases and observe that
the proposed algorithm often achieves scheduling performance far better than the obtained performance {\em lower} bounds. Furthermore, we will only assume the average interference constraint, but it is straightforward to incorporate the instantaneous interference requirement in the solutions.

%a distributed scheduling algorithm with single round contention, called Interference Regulated Distributed Scheduler (IRDS), and obtain its %performance bounds by characterizing $\alpha$. IRDS  has low complexity but lacks exploiting diversity gain of the fading channels. In the next section, we will introduce a distributed scheduler with multiple discrete contention periods that is capable of exploiting the diversity gain.

\subsection{Contention Resolution Phase in CADS}

\begin{figure*}[htp]
\centerline{ \subfloat[Successful Contention ]{\includegraphics[width=3.0in]{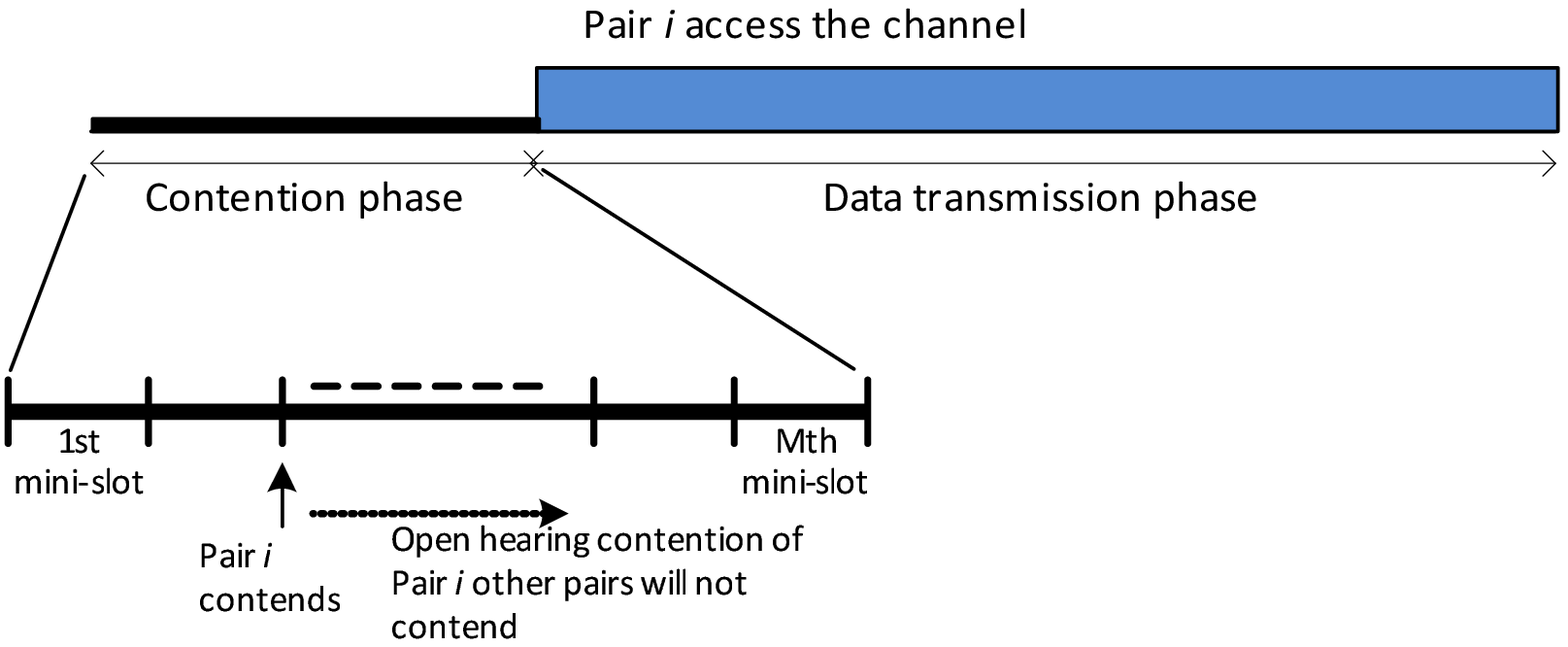}
\label{fig:cont1}} \\ \subfloat[Collision in Contention Phase]{\includegraphics[width=3.0in]{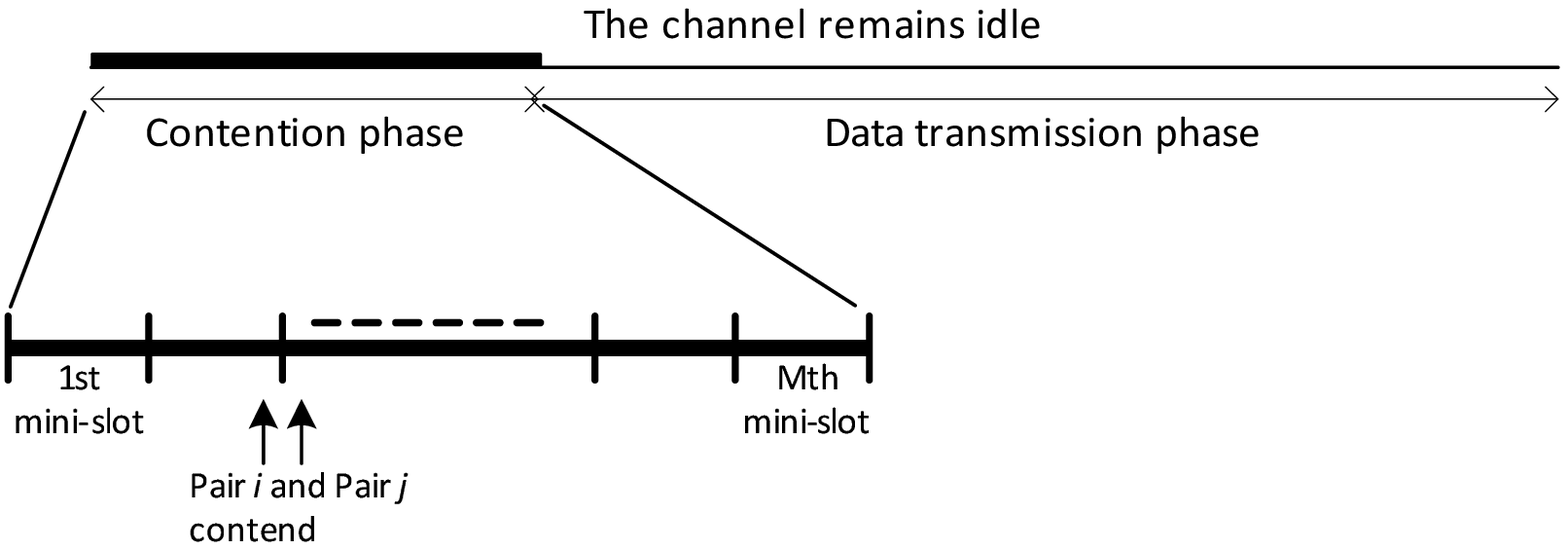}%
\label{fig:cont2}  }} \caption{ Illustration of Algorithm 2}
\end{figure*}

%In the previous section, we showed that SRC  algorithm can not take advantage of diversity gain of wireless fading channels. 
In this part, we introduce the common operational principles shared by all CADS-type distributed schedulers that will be introduced in the subsequent sections. Operation of a CADS takes place in slotted time in two phases: (i) contention phase and (ii) data transmission phase. The contention phase is composed of $M$ mini-slots, each of which is of enough duration to detect contention signals from other devices, i.e., a mini-slot must be at least $8 \mu\mbox{s}$ in IEEE 802.11b. If $\tau$ is the ratio of the mini-slot duration to the duration of a regular time slot, then the parameter $M\tau$, which will appear in our derivations below, signifies the fraction of time spent to resolve collisions by means of implementing a contention resolution phase before the transmission of actual data.   

The contention from devices for the time slot $t$ is resolved as follows. The $i$th D2D pair selects a mini-slot $m \in \brparen{1, \ldots, M+1}$ to send its contention signal.  The selected mini-slot $m$ depends on the pair $i$'s weight $W_i(t)$ that incorporates queue backlog, direct channel and interference channel information into a single parameter. If pair $i$ senses a contention signal from another pair before the $m$th mini-slot, it stops contending for the channel and defers its data transmission to the next time slot, i.e., ${\cal I}_i(t) = 0$. Otherwise, it sends a contention signal in the beginning of the $m$th mini-slot. If no collision is sensed, the $i$th D2D pair obtains the access for the channel to transmit its data for the remaining part of the time slot, called the data transmission phase, commencing after the contention phase. If a collision is sensed, then the time slot remains idle and no data transmission takes place. These steps are visually illustrated in Fig. 4 and summarized in Algorithm 1 below.

Secondly, we assume that when a D2D pair detects a contention from other devices, it stops contenting the channel, i.e., the first pair which contends for the channel, obtains the right to access the channel and start transmitting in data transmission phase as illustrated in Fig. \ref{fig:cont1}. Furthermore, if more than one user contend in the same mini-slot, we say that collision takes place and the channel remains idle in data transmission phase as illustrated in Fig. \ref{fig:cont2} to prevent redundant interference to AP. The basic operational steps of CADSs is summarized in Algorithm \ref{alg:CSMA_Based}.

\begin{algorithm}
\caption{Contention Resolution in CADS}  
	1. At the beginning of each time slot $t$, the pair $i$ picks a mini-slot $m \in \{ 1, \ldots M \}$ based on its weight $W_i(t)$. %(summarizing the queue backlog, direct channel gain and interference channel gains information) and keep sensing the channel until the $m$th %mini-slots.
	
	2.  If a contention signal is heard before the $m$th mini-slots, then the pair $i$ suspends its contention, it will set ${\cal I}_i(t) = 0$. %and remains silent in the data %transmission phase of time slot $t$, i.e., 
	
	3. If no contention signal is heard, the pair $i$ transmits a contention signal at the beginning of the $m$the mini-slot.
	
		- If a collision is detected, the pair $i$ sets ${\cal I}_i(t) = 0$. %i.e., another pair contends for the channel in the $m$th mini-slot, the pair $i$ remains silent during the %time slot $t$ (i.e., ). 
		
		- If no collision is detected, the pair $i$ sets ${\cal I}_i(t) = 1$, and starts its transmission at the beginning of data transmission phase.
	
	4. %If ${\cal I}_i(t) = 1$, the pair $i$ starts transmitting its data at the beginning of data transmission  phase pertaining to time slot $t$. 
	The whole process restarts in the next time slot. 
	\label{alg:CSMA_Based} 
\end{algorithm} 

Based on the contention resolution algorithm in CADS described above, the D2D pair with the smallest backoff time earns the access rights for the channel. Hence, it is of critical importance to design an efficient association policy mapping smaller backoff times to the larger weights $W_i(t)$ to ensure high utility and to exploit multiuser diversity. Our aim below is to investigate the structure of such efficient policies associating device weights with the mini-slot indices. 

\begin{definition} A mini-slot association policy $\vecbold{\Theta}\paren{\vecbold{x}} = \sqparen{\Theta_1\paren{x_1}, \ldots, \Theta_N\paren{x_N}}$ is a mapping $\vecbold{\Theta}: \Rp^N \mapsto \brparen{1, \ldots, M+1}^N$ such that its $i$th component function $\Theta_i\paren{x_i}$ determines the mini-slot index to which the $i$th pair is assigned given that $W_i(t) = x_i$. 

Further, $\vecbold{\Theta}$ is said to be a threshold policy if all of its component functions can be written as %, where 

\small
\vspace{-0.15in}
\begin{align*}
\Theta_t^{(i)}(x) = \begin{cases} m, & \text{ if } a_m^{(i)}(t) \leq x < a_{m-1}^{(i)}(t), \ \text{ for } m \in \{1,\ldots,M \} \\
M+1, & \text{ if } W_i(t) < 0,
\end{cases}
\end{align*}
\normalsize
where $\Theta_t^{(i)}(x) = M+1 $ indicates that the pair $i$ does not contend for the channel in time slot $t$. 

Note that the input of mapping function is the weight of devices, $W_i(t)$. Hence, if $W_i(t) < 0$, the transmission of the pair $i$ causes more harms to the AP by injecting excessive interference compared to the benefit of the data transmission. %Furthermore, since $W_i(t)$ is %lower bounded by zero and upper bounded by infinity, $a_M^{(i)}(t) = 0$ and $a_0^{(i)}(t) = \infty$ in \eqref{eq:mapping} for all $i$ and %time slot $t$. 
\end{definition}

%$\Theta(W_i(t)) = m_i(t)$. 

Below, we design a threshold-based mini-slot association policy in which the goal is to operate in close proximity of optimal point without imposing high complexity, and to provide fairness between D2D pairs. Furthermore, in numerical section, we compare the performance of the designed policy with 
different threshold-based mini-slot association policies that mainly differ on determining the threshold values $a_m^{(i)}(t)$.

\subsection{CADS with Uniform Mapping}

In CADS with uniform mapping, each pair is assigned to a mini-slot uniformly at random over all available mini-slots. This is achieved by utilizing the distribution of weights $W_i(t)$ of pairs as follows. \footnote{In our model, the fading process is stationary and independent from slot to slot. Hence, the D2D pairs know their instantaneous channel gains, and can learn channel distributions by observing the channel over a period of time \cite{Vaart98}.} Let $F_{i,t}(x)$ is the conditional cumulative distribution function (CDF) of $W_i(t)$ at time slot $t$ given that $W_i(t)$ is larger than zero, and let $f_{i,t}(x)$ be the corresponding probability density function (PDF). Given $Q_i(t)$ and $Z(t)$, each device can obtain $F_{i,t}(x)$ as follows:
\begin{align}
F_{i,t}(x) = \Prob{W_i(t) \leq x | W_i(t) > 0, Q_i(t), Z(t)}
\end{align}

Furthermore, let $(F_{i,t})^{-1}(\cdot)$ be the inverse function of $F_{i,t}(\cdot)$. The following lemma indicates how to select the threshold values $a_m^{(i)}(t)$ to achieve uniform distribution over all mini-slot indices.

%Each pair determines the mini-slot in which it will contend by comparing its weight, $W_i(t)$ by the threshold values as illustrated in Fig. \ref{fig:sel_dist}.

\begin{lemma}
	%If $W_i(t) < 0 $, SU $i$ sets ${\cal I}_i(t) = 0$ and do not contend for the channel. 
	%Else it determines the mini-slot as: \\
	In CADS with uniform mapping, the mapping function, $\boldsymbol{\Theta}_t$, is as follows:

\vspace{-0.15in}
			\begin{align*}
\Theta_t^{(i)}(x) = \begin{cases} m, & \text{ if } (F_{i,t})^{-1}\left( \frac{M-m}{M}\right) \leq x < (F_{i,t})^{-1}\left( \frac{M-m-1}{M}\right), \\ 
&\ \ \ \ \ \ \ \ \ \ \ \ \ \ \ \ \ \ \ \ \text{ for } m \in \{1,\ldots,M \} \\
M+1, & \text{ if } W_i(t) < 0,
\end{cases}
\end{align*}
	\label{def:mapping_dist}
\end{lemma}

The mini-slot association policy above ensures that each pair picks a mini-slot uniformly at random, i.e., with a probability $\frac{1}{M}$ given that its weight is positive. The goal here is to minimize the probability collisions by spreading the contention probabilities uniformly over all mini-slots. Furthermore, this mini-slot association policy also enforces the scheduling of schedule a good D2D pair with respect to the current channel and queue states. However,  it should be noted that such a uniform mapping policy, although promising, does not necessarily guarantee the scheduling of the best user, i.e., the user that has the maximum weight, as discussed subsequently.

%The algorithm can be summarized as follows.

%Collisions in contention phase aims to prevent abundant interference to the PBS during the whole slot. The main question here is that how SUs decide in which mini-slot they will contend based on their %weights in that slot.

% \begin{figure}
%\includegraphics[width=3.5in]{selection_fig.eps}
%   \caption{Illustration of mapping in Rayleigh fading channels }
%	\label{fig:sel_dist}
%\end{figure}

\subsection{Performance Analysis of CADS with Uniform Mapping}

%In this subsection, we show the performance analysis of the algorithm under the assumption that the direct and interference channel gains are %iid. 
Here, we characterize the performance of CADS with uniform mapping by studying the performance loss due to fraction of time allocated to the contention resolution period, collisions in the contention period and imperfect scheduling.

1) \textit{ Performance Loss Due to Contention Period}: We assume that the length of mini-slots is not negligible. Hence, the devices that are scheduled to transmit, can only use $1-M\tau$ fraction of a whole time slot. That is the loss due to a contention resolution phase with multiple mini-slots, is $M\tau$.

2) \textit{Performance Loss Due to Collisions in the Contention Phase}: The CADS with uniform mapping cannot prevent collision in the contention phase perfectly. Whenever such a collision occurs, all D2D pairs remain silent during the data transmission phase, and the channel becomes under-utilized. In the sequel, we characterize this loss.

3) \textit{Performance Loss Due to Imperfect Scheduling}: The CADS with uniform mapping does not always schedule the D2D pair that has the maximum weight. The main underlying reason behind this phenomenon is that each device is assigned to a mini-slot uniformly at random with respect to their respective weights. Although this provides fairness among devices in giving the access rights to the channel (i.e., devices with lower and higher weights are treated equally), it can lead to assignment of channel access rights to the devices with smaller weights. Let us define the expected loss due to such an imperfect scheduling as $\beta$, where $0 \leq \beta \leq 1$.

We do not try to characterize $\beta$ analytically for arbitrary number of devices and mini-slots due to intractability of calculations. Hence, $\beta$ will appear as a parameter in Theorem 3 determining the performance of the CADS with uniform mapping. Our simulation results indicate that the loss due to imperfect scheduling is negligibly small for i.i.d. channels.

We need the following lemma to characterize the performance of the CADS with uniform mapping. 

 \begin{lemma}
\label{lemma:bound}
 	The  CADS with uniform mapping satisfies the following inequality:
	\small
 	\begin{align*}
 		\E{\sum_{i=1}^N{\cal I}_i(t)W_i(t)} & \geq \E{W^*(t)}\alpha.
 	\end{align*}
	\normalsize
	where $\alpha = (1-M\tau)(1-\beta)\left(\frac{N}{M} \sum_{k=1}^M\frac{N(M-k)^{N-1}}{(M-k+1)^N - (M-k)^N} \right)$
 \end{lemma}
 
\begin{IEEEproof} The proof is given in Appendix \ref{proof:bound_weight}. \end{IEEEproof} 
		
%\endproof
Lemma \ref{lemma:bound} indicates that the sum of average weights achieved by the uniform mini-slot association policy is larger than a fraction of the maximum weight. By using Lemma \ref{lemma:bound}, we next characterize the performance of the CADS with uniform mapping.

 \begin{theorem}
\label{thm:optimalcontrol_dist_sel}
 Suppose $\boldsymbol{x^*} = [x_1^*, \ldots, x_N^*]$ is the average arrival rates produced the CADS with uniform mapping. Then, for any flow parameter  $V > 0$, the algorithm achieves the following performance bound:
 \begin{align*}
    \sum_{i=1}^N  U_i(x_i^*) &\geq \alpha U^* - \frac{B_2}{V} 
		\end{align*}
while bounding the long-term expected queue lengths as:	
		\begin{align*}
    \limsup_{T \rightarrow \infty} \frac{1}{T} \sum_{\tau =
    0}^{T-1}\sum_{i=1}^N \E{Q_i(\tau)} &\leq
    \frac{B_2+V\kappa}{\epsilon_3},
 \end{align*}
 where $B_2,\epsilon_3, \kappa > 0$ are constants, and $U^*$ is the optimal
 aggregate utility, i.e., the solution of the problem in
 \eqref{eq:opt-objective}-\eqref{eq:const-stability}. %and
 %$\bar{U}$ is the maximum possible aggregate utility.

%This theorem shows that it is possible to get arbitrarily close to
%the optimal utility by choosing $V$ sufficiently large at the
%expense of proportionally increased average queue sizes.
\end{theorem}

\begin{IEEEproof} The proof follows from Corollary 5.2 in \cite{Georgiadis} in the light of information given in Lemma \ref{lemma:bound}, and given in Appendix \ref{proof:interference_regulated_sel}. \end{IEEEproof} %The proof is given in Appendix \ref{proof:interference_regulated_sel}. \end{IEEEproof} 

%\textit{Remark:} %Assuming that $\tau$ is negligibly small, 

  }
{\allowdisplaybreaks\section{Numerical Results}

\subsection{Simulation Setup and Parameters}

In our numerical experiments, we consider a two-tier network, in which D2D pairs are communication with each other while causing interference to CBS. Furthermore, we consider logarithmic utility function for all D2D pairs \footnote{We utilize logarithmic utility function to provide proportional fairness.}. Specifically, pair $i$ obtains utility of $\log(1+x_i)$ at the rate of $x_i$. The rates depicted in the graphs are the sum of arrival rates of all users and the unit of the plotted rates is natts/channel use. First, the main channel between D2D pairs and interference
channel between transmitters of pairs are modeled as i.i.d.
Rayleigh fading Gaussian channels. Thus, the main and
interference power gains are exponentially distributed with
means 2 and 1, respectively. The noise normalized power is $P=1$. Furthermore, in experiments, we only consider average interference constraint.
We compare the performance of our algorithm with different mini-slot association policies and that of widely used regulated queue approach.
\subsubsection{CADS with Optimal Weight Mapping}

%In previous subsection, we designed distributed algorithm with uniform mapping, which minimizes collision instances in each mini-slot. %However, in our problem, minimizing collisions does not correspond to optimized performance. This is because, collisions in the first %mini-slots corresponds to larger loss in the performance compared to collisions in later mini-slots, since in the first mini-slot, only pairs %with larger value of $W_i(t)$ contend. Hence, 
In this subsection, we design a mapping function  such that it maximizes the expected weight of D2D pairs by neglecting the effect of the imperfect scheduling. That is to say, the sequence of threshold value is determined as the solution of the following optimization problem:

%\small
\vspace{-0.15in}
\begin{align*}
 \max_{a_m} \ &\E{{\cal I}_i(t) W_i(t)} = \sum_{m=1}^M \E{{\cal I}_i(t) W_i(t)| \Theta(W_i(t)) = m} %\\
%&= \sum_{m=1}^M \E{W_i(t)| \Theta(W_i(t)) = m}\E{{\cal I}_i(t)| \Theta(W_i(t)) = m}  \\
%&  = \int_{a_1}^\infty \left( x f(x) dx \right)\left(1-F(a_1)\right)^{N-1} (1-F(a_1)) \\
%&+\int_{a_2}^{a_1} \left( x f(x) dx \right)\left(1- F(a_2)\right)^{N-1}\left( F(a_1) - F(a_2) \right)\\
%&+\ldots \\
%&+\int_{0}^{a_{M-1}} \left( x f(x) dx \right) \left( F(a_{M-1}) \right)
\end{align*}

%\normalsize

Since the above optimization problem is highly non-linear and dependent on distribution of fading process, we are not able to obtain closed-form solution. Hence, we numerically solve the problem in the numerical result section. Furthermore, obtaining the performance bound on this mapping function is non-trivial due to having no access to closed-form solution. However, we should note that since the algorithm maximizes the sum of weights, it results in better bounds obtained for the one with uniform mapping given in Lemma \ref{lemma:bound} and Theorem \ref{thm:optimalcontrol_dist_sel}. Hence, it results in better performance compared to uniform mapping at the expense of high complexity.

\subsubsection{CADS with Linear Mapping}

%In the previous subsections, we assume that each D2D pairs obtains the distribution function and based on this distribution function maps its %weight in slot $t$ into the number of mini-slot in which it contends. This requires the knowledge of distribution of its channel gains, and %calculating mapping function. For some of the nodes with limited power and memory, it may not possible to calculate this mapping function, %e.g., wireless sensors. 

In this section, we explicitly consider this practical challenge and propose an easily implementable and efficient algorithm for the users having limited power and memory. This time, we utilize a liner mapping discrete mapping function $\Theta$, is defined as follows:. 
%\begin{definition}
%In the distributed algorithm with linear mapping, the mapping function, $\Theta$, is defined as follows:
			\begin{align*}
\Theta(W_i(t)) = \begin{cases} m, & \text{ if } \frac{(M-m-1)W_{max}}{M}  \leq W_i(t) < \frac{(M-m)W_{max}}{M}, \\ 
&\ \ \ \ \ \ \ \ \ \ \ \ \ \ \ \ \ \ \ \ \text{ for } m \in \{1,\ldots,M \} \\
M+1, & \text{ if } W_i(t) < 0,
\end{cases}
%\label{eq:mapping}
\end{align*}
%	\label{def:mapping_dist}
where $W_{max}$ is ideally a value that the the number of realization of $W_i(t)$ which are larger than $W_{max}$, is small. However, since we assume that the users are not capable of calculating the mapping according the distribution function, they agree a value for $W_{max}$.
%\end{definition}
 
%We should note that the performance of the algorithm with linear mapping mainly depends on two factors: (1) The selection of $W_{max}$. That %is to say, if $W_{max}$ is too small, then the contentions performed by D2D pairs, get clustered around the first mini-slots. This will %result high volume of collisions which decreases the performance of the algorithm. On the contrary, if it is too large, then users tend to %select later mini-slots to contend, and this again causes high volume of collisions and degrades the performance of the algorithm. Thus the %selection of $W_{max}$ plays the crucial role for the performance of the algorithm. (2) The shape of actual distribution function of the %weights. Since we assume linear mapping, if the shape of the distribution is closer to a linear function (or the error between actual %function and linear approximation of the function is small), then we can say that the algorithm performs well. Note that, any other mapping %function can be used instead of linear mapping based on the shape of the distribution function. However, linear function is simple %approximation and does not require high power or memory for calculation. Furthermore, it works well for a wide range of distribution %function, e.g., Rayleigh or Nakagami fading.  
\subsubsection{Interference Regulated Distributed Scheduler}
 In the sequel, we modify the regulated queue model to consider the interference constraint, and it is called Interference Regulated Distributed Scheduler (IRDS), which is based on the baseline algorithm in \cite{Xue13}. 

The operation of IRDS again has again two phases: (i) contention phase and (ii) data transmission phase. To facilitate the discussion, we introduce two new random variables related to contention and scheduling phases. The first one is the contention variable, $a_i(t)$, that is 1 with probability $\frac{1}{N}$, and 0 with probability $\frac{N-1}{N}$. 
%has a probability distribution given by
%\begin{align}
%a_i(t)=\begin{cases} 1, & \text{ w.p.  } \frac{1}{N} \\
% 0, & \text{ w.p.  } \frac{N-1}{N}.
%\end{cases}
%\end{align}
The second one is the transmission variable, $p_i(t)$, that is 1 with probability  $\frac{e^{W_i(t)}}{e^{W_i(t)}+1}$, and 0 with probability $\frac{1}{e^{W_i(t)}+1}$, %independent over pairs, and has distribution given by
%\begin{align}
%p_i(t)=\begin{cases} 1, & \text{ w.p.  } \frac{e^{W_i(t)}}{e^{W_i(t)}+1} \\
% 0, & \text{ w.p.  }  \frac{1}{e^{W_i(t)}+1},
%\end{cases}
%\end{align}
where $W_i(t)$ is the weight of D2D pair $i$ defined in \eqref{eq:weight}. %The dynamics of
%real queue, $Q_i(t)$, and virtual queue, $Z(t)$ are as in
%\eqref{eq:real_queue} and \eqref{eq:interference_queue}. 
We note that the transmission variable takes also into account interference level that is caused to AP.  This is the reason why, the algorithm is regulated with respect to the interference level.

The scheduling decision of D2D pair $i$ depends on the following three conditions: 

Condition (1): The contention of pair $i$  is successful, i.e.,
$a_i(t)\prod_{j \neq i}(1-a_j(t))= 1$.

Condition (2): None of the neighboring pairs were scheduled in the
previous time slot, i.e., $\sum_{j \neq i} {\cal I}_j(t-1) = 0$

Condition (3): The transmission variable $p_i(t) = 1$. 
 
Based on these three conditions, the scheduling phase consists of three different cases, as given by Algorithm \ref{alg:int_reg}. 
\begin{algorithm}
\caption{Scheduling Phase of IRDS}   
    Each D2D pair  schedules ${\cal I}_i(t)$ according to 

Case 1: ${\cal I}_i(t) = 1$ if Conditions (1)(2)(3) holds.

Case 2: If Condition (1) does not hold and Condition (3) holds, then
${\cal I}_i(t) = {\cal I}_i(t-1)$.

Case 3: Otherwise, ${\cal I}_i(t) = 0$.

\label{alg:int_reg}
\end{algorithm}

Notice that IRDS only consider single round contention unlike the proposed CADSs, which limits the performance of the algorithm as shown in simulation results.

 \begin{figure}
\centerline{{\includegraphics[width=3.0in]{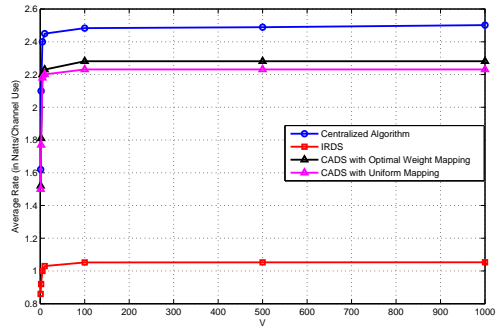}
\label{fig:D2D_V}}}   \caption{Performance Analysis of Algorithms with respect to $V$ }
\end{figure}

\subsection{Simulation Results}

In Fig. 5, we investigate the effect of system parameter $V$
in our dynamic control algorithms. We take interference constraint, $\gamma = 1$ and the ratio of length of a mini-slot to the length of a slot is taken as $0.0001$, i.e., $\tau = 0.0001$. Furthermore, the number of mini-slots, $M$, is 200. As expected, the average rate of all algorithms increase with increasing $V$
and
Fig. 5 shows that the long-term utilities for $V \geq$ converges to their optimal values fairly
closely verifying the results in Theorem 1. Furthermore, the distributed algorithm with the best performance is CADS with optimal weight mapping achieving over $90 \%$ of the average rate of centralized algorithm. CADS with uniform mapping exhibits a performance fairly close to the one with optimal weight mapping. However, IRDS is the distributed scheduler with worst performance achieving only approximately $40 \%$ of the average rate of centralized algorithm.

\begin{figure*}
\centerline{ \subfloat[Performance with respect to $\gamma$,
$\gamma$]{\includegraphics[width=3.0in]{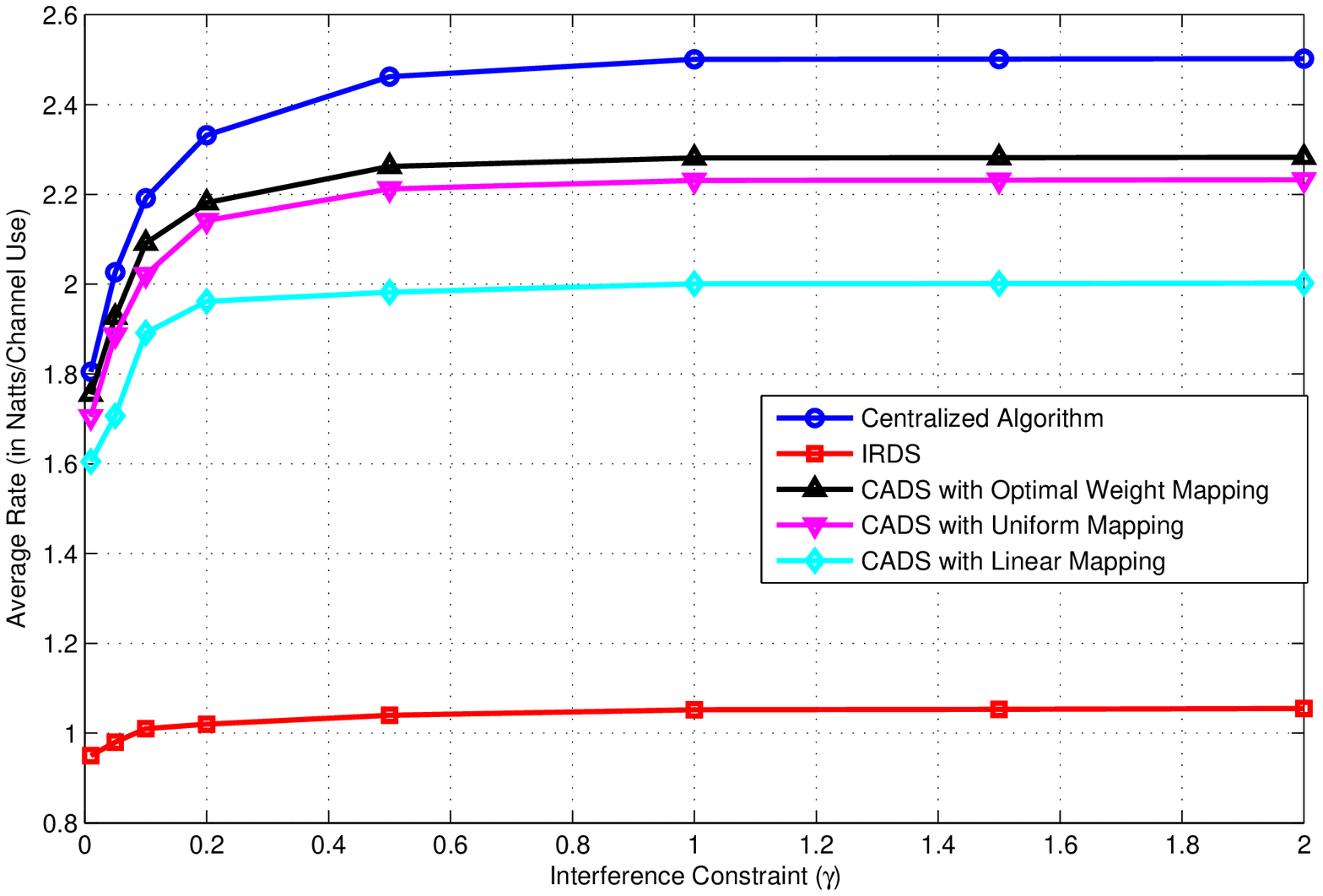}
\label{fig:D2D_gamma}} \subfloat[Performance with respect to $N$,
$\E{g_2}$]{\includegraphics[width=3.0in]{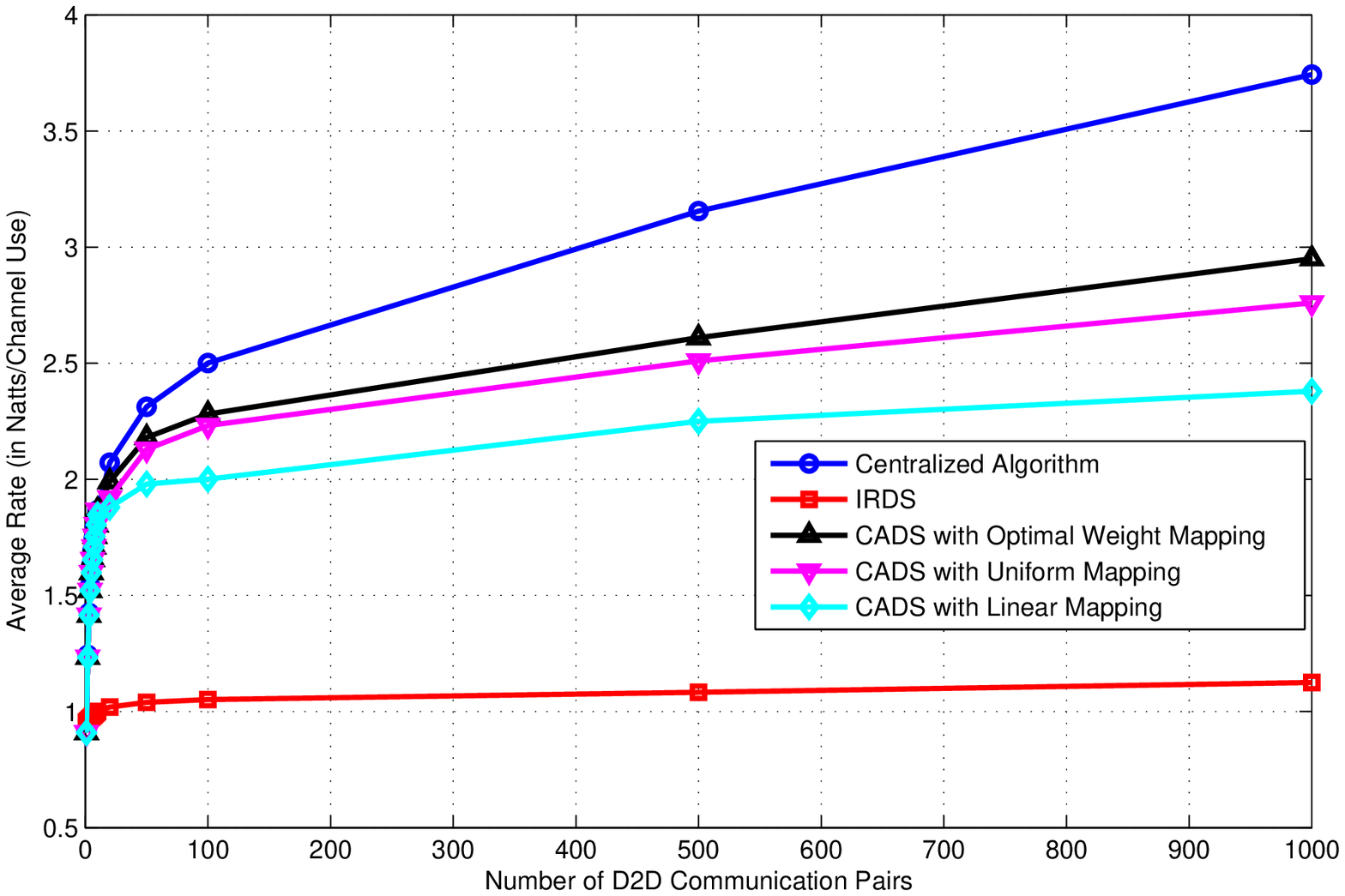}%
\label{fig:numnode}  }} \caption{Performance Analysis of Algorithms with respect to $\gamma$ and $N$ }
\end{figure*}

For the rest of the experiments, we take $V= 200$. In Fig. \ref{fig:D2D_gamma} and \ref{fig:numnode}, we anaylze the effect of interference constraint, $\gamma$, number of D2D pairs, $N$, on the system performance, respectively. As illustrated in Fig. \ref{fig:D2D_gamma}, the average rate for all algorithms increases with increasing $\gamma$. This is because for low $\gamma$ values, in order to satisfy a tight interference constraint, a larger fraction of time-slots remains idle, i.e., smaller number of transmission opportunities are given to D2D pairs.  Starting around
$\gamma$ = 1, the interference constraint becomes inactive, since the constraint is realized with strict inequality. From Fig \ref{fig:numnode}, we first notice that the performance of IRDS does not change with the increasing number of pairs and only achieves $\% 90$ of the rate achieved by the centralized algorithm when the number of pairs is one. Thus, we can conclude that IRDS cannot take advantage of diversity gain of fading channels. On the other hand, CADS with optimal weight, uniform and linear mapping achieves a performance that is closer to the centralized algorithm. As expected, CADS with optimal weight mapping performs the best whereas the one with linear mapping has the worst performance. In linear mapping, we sacrifice some performance in favor of reducing the complexity of mapping. Furthermore, as the number of pairs increases, the collisions during contention phase increases. This results in an increase in the difference between the performance of centralized and CADSs.

\begin{figure*}
\centerline{ \subfloat[$\tau = 0.0001$,
$\gamma$]{\includegraphics[width=3.0in]{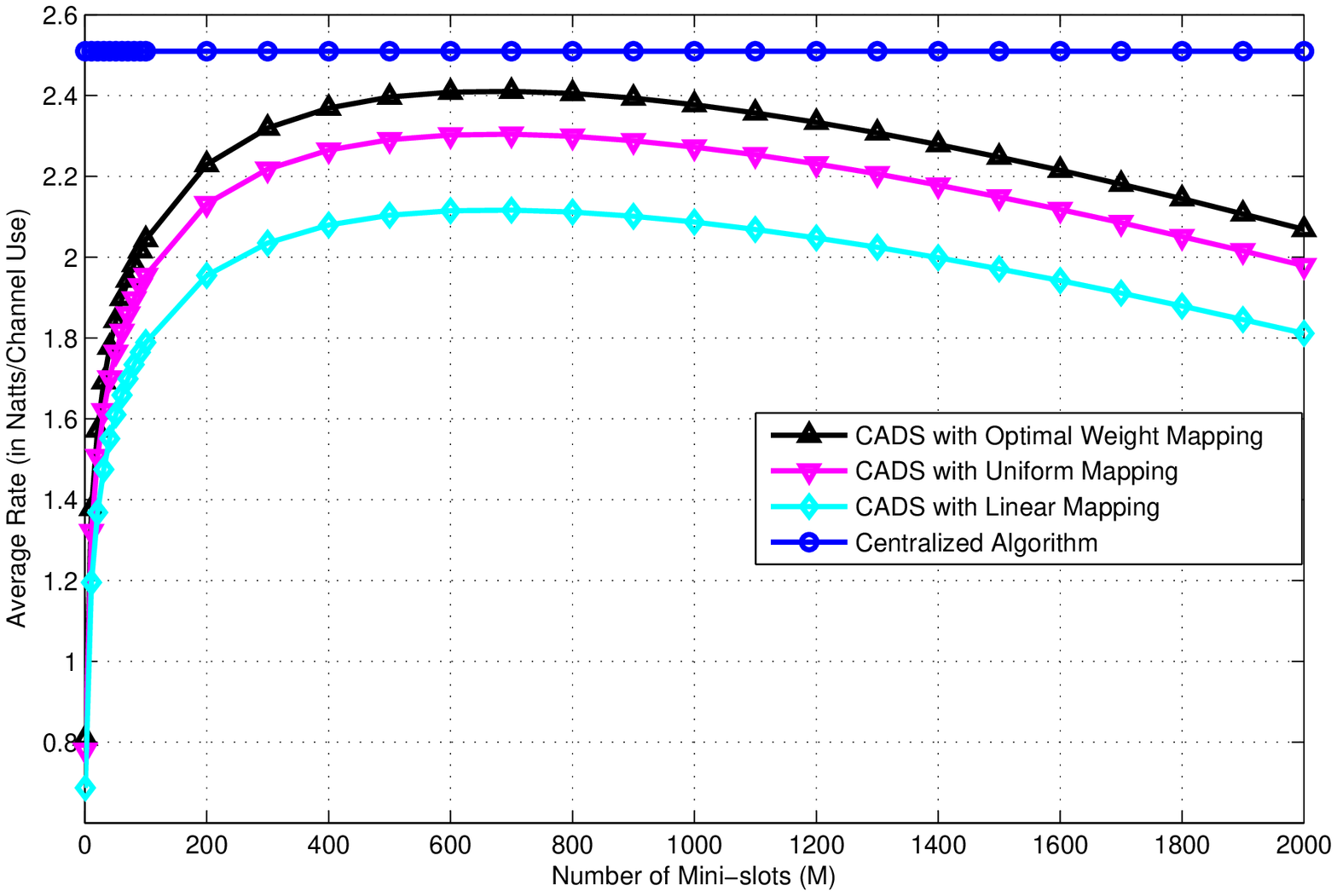}
\label{fig:M1}} \subfloat[$\tau = 0.0002$,
$\E{g_2}$]{\includegraphics[width=3.0in]{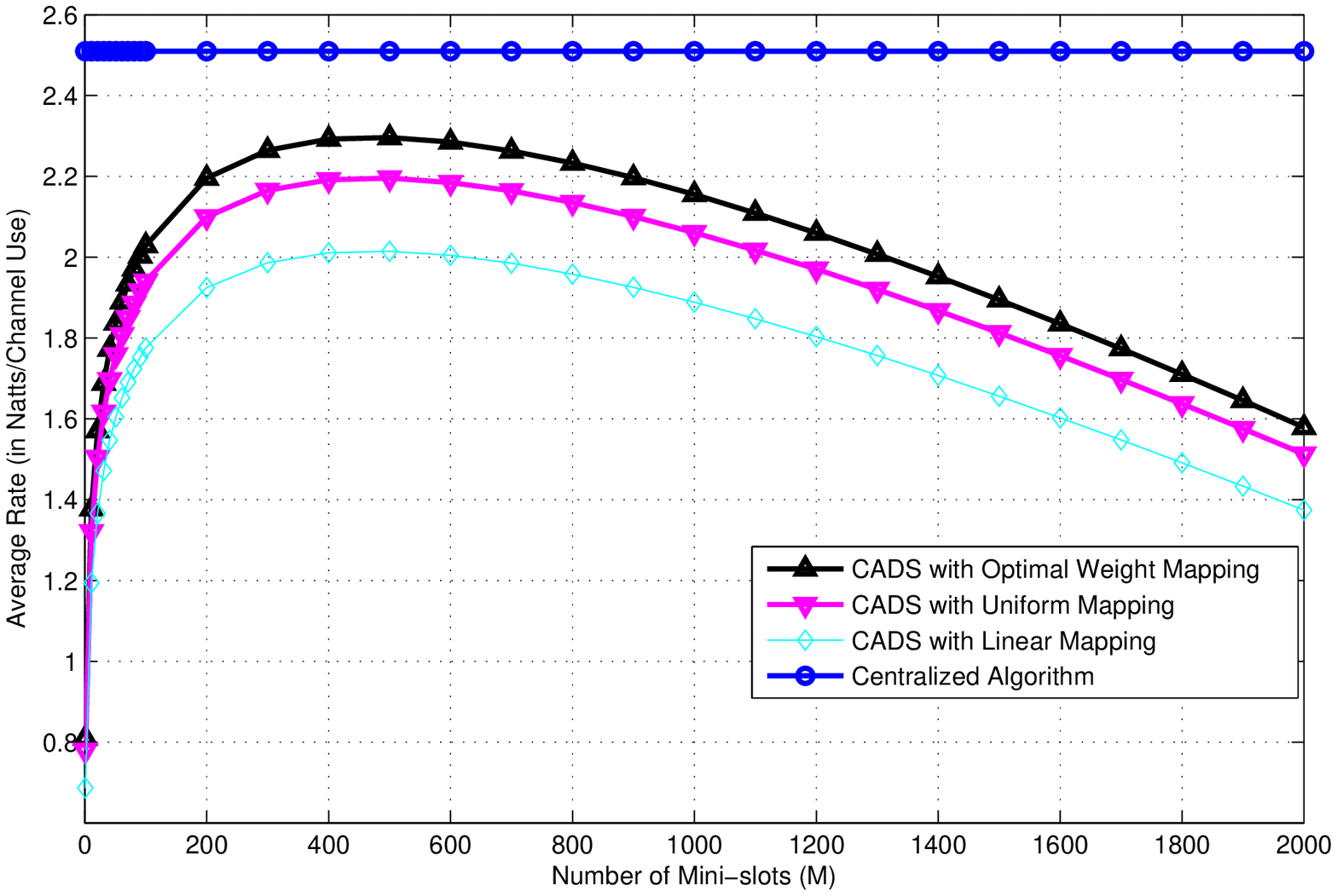}%
\label{fig:M2}  }} \caption{Performance Analysis of CSMA-based distributed algorithms with respect to the ratio of length of mini-slots to the one of slot.}
\end{figure*}

Next, we analyse the effect of the number of mini-slots on the performance of CADSs. We take the number of SUs as 100. In Fig. \ref{fig:M1}, we take $\tau = 0.0001$ and in Fig. \ref{fig:M2}, we take $\tau = 0.002$. As illustrated in Fig. \ref{fig:M1}, the average data rate increases
initially with increasing $M$. This is due to fact with increasing $M$, the network experiences less collision. However,
when $M$ is high, the emphasis on reducing collisions becomes less significant, and the loss due to length of contention phase, i.e., $M\tau$ increases. As a result, the performance of CADSs gets worse. The optimal $M$ is achieved when $M$ is around 600. In Fig. \ref{fig:M2}, we notice that the optimal $M$ is around 400 and the decrease of the average rate due to having long contention phase is sharper compared to the case when $\tau = 0.0001$.

\begin{figure*}
\centerline{ \subfloat[Performance with respect to $\gamma$,
$\gamma$]{\includegraphics[width=3.0in]{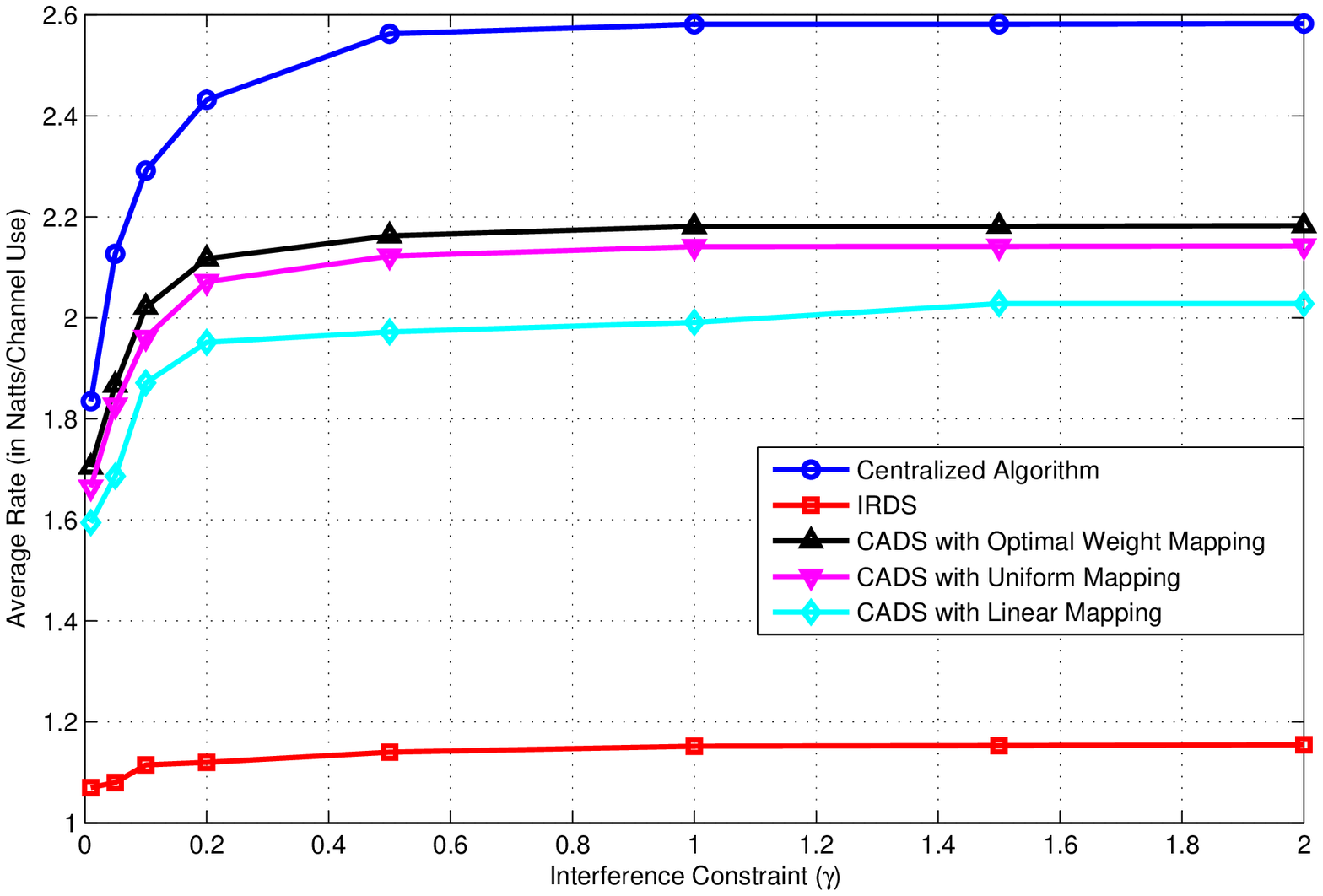}
\label{fig:gamma_NU}} \subfloat[Performance with respect to $\gamma$,
$\E{g_2}$]{\includegraphics[width=3.0in]{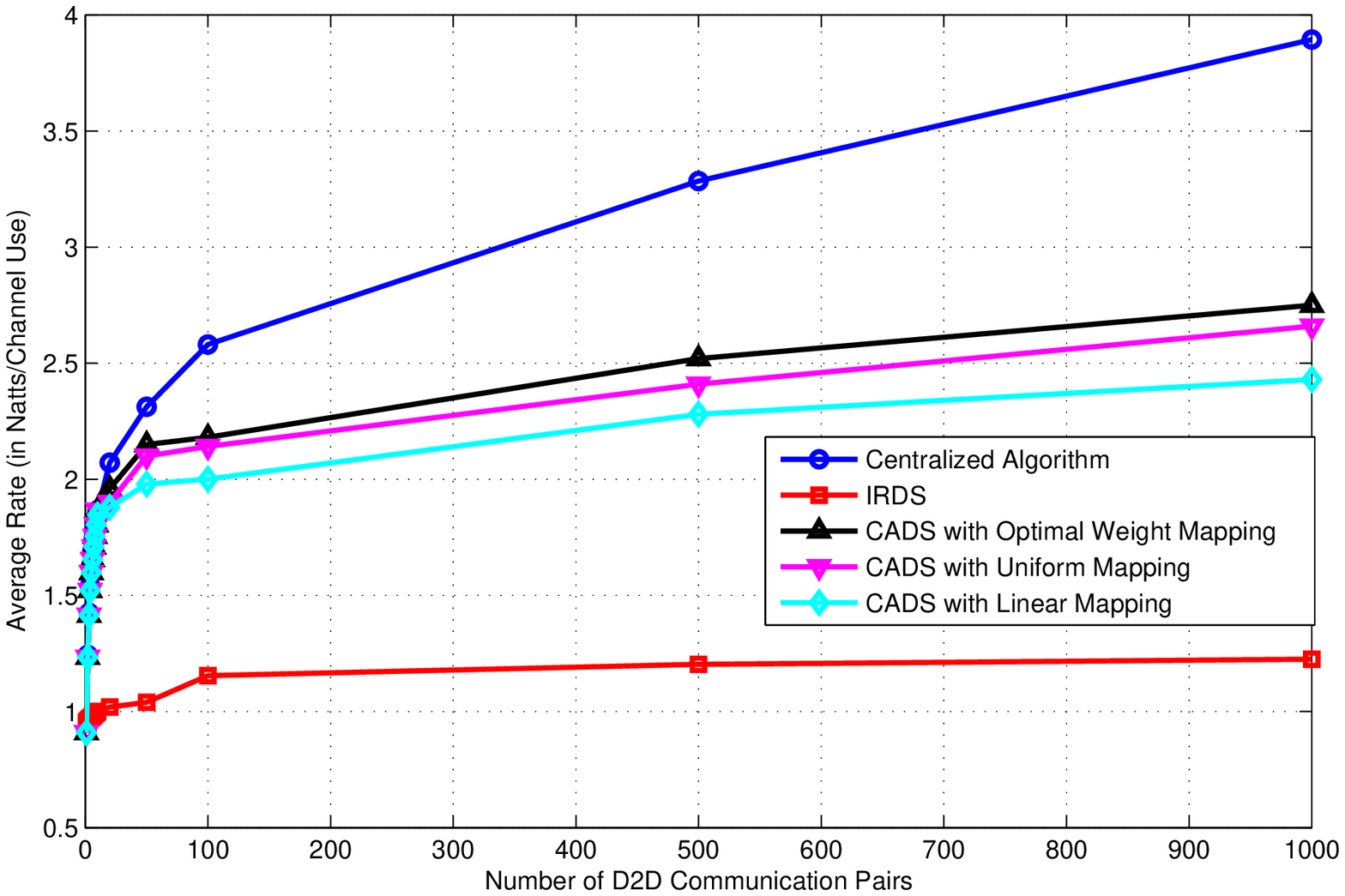}%
\label{fig:numnode_NU}  }} \caption{Performance Analysis of Algorithms with respect to $\gamma$ and $N$ with non-iid channels}
\end{figure*}

Lastly, we investigate the performance of algorithms with respect to $\gamma$ and number of D2D pairs, $N$, when the channels are non-iid. Precisely, the main and interference channel gains of D2D pairs are chosen at
random, uniformly distributed in the intervals [1.2, 2.8], and [0.2,1.8], respectively. In addition, we take 10 runs and the rates depicted in Fig. \ref{fig:gamma_NU} and \ref{fig:numnode_NU} are the average of these 10 runs. From Fig. \ref{fig:gamma_NU} and \ref{fig:numnode_NU}, we notice that CADS with optimal weight and uniform mappings perform slightly worse compared to the case when the channel gains are iid. %, e.g., %the algorithm with optimal weight mapping now achieves approximately $85 \%$ of centralized algorithm whereas it achieves $90 \%$ of %centralized algorithm when the channel gains are iid. 
The reason is that when the channel gains are non-iid, the algorithms do not schedule the pair with the highest weight in high number of instances frequency due to having different mapping interval for each pair. }
{\allowdisplaybreaks\section{Conclusion}

In this paper, we considered the problem of resource allocation in
wireless two-tier network where D2D communication pairs have information to be shared.  However, the interference to the CBS should be kept arbitrarily low caused by the transmission between D2D pairs. First, we studied the stability region of such interference-aware network, and compare the region to the case when there is no interference constraint. Then, we
described a cross-layer dynamic algorithm, and we proved that our algorithm achieves utility
arbitrarily close to achievable optimal utility. 

Finally, we investigate distributed algorithms, where each pair decides to transmit or not according to local information. We design a channel-aware distributed scheduling algorithm, which is a threshold based mini-slot association policy. We derive the performance bound achieved by this algorithm, and compare the performance of the proposed algorithm with widely used regulated queue approach and different mini-slot association policies. Via simulations, we show that the proposed algorithm achieves high performance, and the reduction in average rate in the proposed algorithms due to implementing distributed approach is relatively small.
}
%\tiny
\bibliographystyle{IEEEtran}
\bibliography{d2dcom,IEEEabrv}

% Generated by IEEEtran.bst, version: 1.13 (2008/09/30)
\begin{thebibliography}{10}
\providecommand{\url}[1]{#1}
\csname url@samestyle\endcsname
\providecommand{\newblock}{\relax}
\providecommand{\bibinfo}[2]{#2}
\providecommand{\BIBentrySTDinterwordspacing}{\spaceskip=0pt\relax}
\providecommand{\BIBentryALTinterwordstretchfactor}{4}
\providecommand{\BIBentryALTinterwordspacing}{\spaceskip=\fontdimen2\font plus
\BIBentryALTinterwordstretchfactor\fontdimen3\font minus
  \fontdimen4\font\relax}
\providecommand{\BIBforeignlanguage}[2]{{%
\expandafter\ifx\csname l@#1\endcsname\relax
\typeout{** WARNING: IEEEtran.bst: No hyphenation pattern has been}%
\typeout{** loaded for the language `#1'. Using the pattern for}%
\typeout{** the default language instead.}%
\else
\language=\csname l@#1\endcsname
\fi
#2}}
\providecommand{\BIBdecl}{\relax}
\BIBdecl

\bibitem{Lei12}
L.~Lei, Z.~Zhong, C.~Lin, and X.~Shen, ``{Operator controlled deviceto- device
  communications in lte-advanced networks},'' \emph{IEEE Wireless Commun.
  Mag.}, vol.~19, no.~3, pp. 96--104, Jun. 2012.

\bibitem{Fodor12}
G.~Fodor, E.~Dahlman, G.~Mildh, S.~Parkvall, N.~Reider, G.~Miklos, and
  Z.~Turanyi, ``{Design aspects of network assisted device-to-device
  communications},'' \emph{IEEE Commun. Mag.}, vol.~50, no.~3, pp. 170--177,
  Mar. 2012.

\bibitem{tassiulas}
L.~Tassiulas and A.~Ephremides, ``Jointly optimal routing and scheduling in
  packet ratio networks,'' \emph{IEEE Transactions on Information Theory},
  vol.~38, no.~1, pp. 165 --168, Jan. 1992.

\bibitem{shroff}
X.~Liu, E.~K.~P. Chong, and N.~B. Shroff, ``A framework for opportunistic
  scheduling in wireless networks,'' \emph{Computer Networks}, vol.~41, no.~4,
  pp. 451--474, 2003.

\bibitem{urgaonkar}
R.~Urgaonkar and M.~J. Neely, ``Opportunistic scheduling with reliability
  guarantees in cognitive radio networks,'' \emph{IEEE Trans. Mob. Comput.},
  vol.~8, no.~6, pp. 766--777, 2009.

\bibitem{jaramillo}
J.~J. Jaramillo and R.~Srikant, ``Optimal scheduling for fair resource
  allocation in ad hoc networks with elastic and inelastic traffic,'' in
  \emph{INFOCOM}, 2010, pp. 2231--2239.

\bibitem{kelly}
\BIBentryALTinterwordspacing
F.~P. Kelly, A.~K. Maulloo, and D.~K.~H. Tan, ``{Rate Control for Communication
  Networks: Shadow Prices, Proportional Fairness and Stability},'' \emph{The
  Journal of the Operational Research Society}, vol.~49, no.~3, pp. 237--252,
  1998. [Online]. Available: \url{http://dx.doi.org/10.2307/3010473}
\BIBentrySTDinterwordspacing

\bibitem{Kar}
X.~Wang and K.~Kar, ``Cross-layer rate control for end-to-end proportional
  fairness in wireless networks with random access,'' in \emph{MobiHoc}, 2005,
  pp. 157--168.

\bibitem{Georgiadis}
L.~Georgiadis, M.~J. Neely, and L.~Tassiulas, ``Resource allocation and
  cross-layer control in wireless networks,'' \emph{Foundations and Trends in
  Networking}, vol.~1, no.~1, 2006.

\bibitem{Yu09}
C.~H. Yu, O.~Tirkkonen, K.~Doppler, and C.~B. Ribeiro, ``On the performance of
  device-to-device underlay communication with simple power control,'' in
  \emph{in Proc. IEEE Vehicular Technology Conf. (VTC)}, Apr. 2009, pp.
  1--–5.

\bibitem{Yu11}
C.~H. Yu, K.~Doppler, C.~Riberio, and O.~Tirkkonen, ``{Resource Sharing
  Optimization for Device-to-Device Communication Underlaying Cellular
  Communication},'' \emph{IEEE Transactions on Wireless Communications},
  vol.~10, no.~8, pp. 2752--2763, 2011.

\bibitem{Janis09}
P.~Janis, C.~Yu, K.~Doppler, C.~Ribeiro, C.~Wijting, K.~Hugl, O.~Tirkkonen, and
  V.~Koivunen, ``{Device-to-device communication underlaying cellular
  communications system},'' \emph{IEEE Trans. Inf. Theory}, vol.~2, no.~3, pp.
  169--178, Mar. 2009.

\bibitem{Min11}
H.~Min, J.~Lee, S.~Park, and D.~Hong, ``{Capacity enhancement using an
  interference limited area for device-to-device uplink underlaying cellular
  networks},'' \emph{IEEE Trans. Wireless Commun.}, vol.~10, no.~12, pp.
  3995--4000, Dec. 2011.

\bibitem{Xiao11}
X.~Xiao, X.~Tao, and J.~Lu, ``A qos-aware power optimization scheme in ofdma
  systems with integrated device-to-device (d2d) communications,'' in \emph{in
  Proceedings of IEEE VTC-Fall}, 2011, pp. 1--5.

\bibitem{Jung12}
M.~Jung, K.~Hwang, and S.~Choi, ``Joint mode selection and power allocation
  scheme for power-efficient device-to-device (d2d) communication,'' in
  \emph{in Proceedings of IEEE VTC-Spring}, 2012, pp. 1--5.

\bibitem{Hakola10}
S.~Hakola, T.~Chen, J.~Lehtomaki, and T.~Koskela, ``Device-to-device (d2d)
  communication in cellular network-performance analysis of optimum and
  practical communication mode selection,'' in \emph{in Proceedings of IEEE
  WCNC}, 2010, pp. 1--6.

\bibitem{Belleschi11}
M.~Belleschi, G.~Fodor, and A.~Abrardo, ``Performance analysis of a distributed
  resource allocation scheme for d2d communications,'' in \emph{in Proceedings
  of IEEE GLOBECOM Workshops}, 2011, pp. 358--362.

\bibitem{Feng13}
D.~F. et~al., ``{Device-to-Device Communications Underlaying Cellular
  Networks},'' \emph{IEEE Transactions on Communications}, vol.~61, no.~8, pp.
  3541--3551, 2013.

\bibitem{Zhang13}
R.~Zhang, X.~Cheng, L.~Yang, and B.~Jiao, ``Interference-aware graph based
  resource sharing for device-to-device communication underlaying cellular
  communication,'' in \emph{in Proc. IEEE WNC}, 2013, pp. 140--145.

\bibitem{Su13}
L.~Su, Y.~Ji, P.~Wang, and F.~Liu, ``Resource allocation using particle swarm
  optimization for d2d communication underlay of cellular networks,'' in
  \emph{in Proceedings of IEEE WCNC}, 2013, pp. 129--133.

\bibitem{Han12}
B.~G.~K. M.~H.~Han and J.~W. Lee, ``Subchannel and transmission mode scheduling
  for d2d communication in ofdma networks,'' in \emph{in Proceedings of IEEE
  VTC-Fall}, 2012, pp. 1--5.

\bibitem{Le12}
L.~B. Le, ``Fair resource allocation for device-to-device communications in
  wireless cellular networks,'' in \emph{in Proceedings of IEEE GLOBECOM},
  2012, pp. 5451--–5456.

\bibitem{Wang14}
Q.~W. et~al., ``{Quality-optimized joint source selection and power control for
  wireless multimedia D2D communication using stackelberg game},'' \emph{IEEE
  Transactions on Vehicular Technology}, vol.~PP, no.~99, pp. 1--1, Sep. 2014.

\bibitem{Xu13}
C.~X. et~al., ``{Efficiency resource allocation for device-to-device underlay
  communication systems: A reverse iterative combinatorial auction based
  approach},'' \emph{IEEE Journal on Selected Areas in Communications},
  vol.~31, no.~9, pp. 348–--358, Sept. 2013.

\bibitem{Li14}
Y.~Li, D.~Jin, J.~Yuan, and Z.~Han, ``{Coalitional games for resource
  allocation in the device-to-device uplink underlaying cellular networks},''
  \emph{IEEE Transactions on Wireless Communications}, vol.~13, no.~7, pp.
  3965–--3977, July 2014.

\bibitem{Chen14}
H.~Chen, D.~Wu, and Y.~Cai, ``{Coalition formation game for green resource
  management in D2D communications},'' \emph{IEEE Communications Letterss},
  vol.~18, no.~8, pp. 1395--–1398, Aug. 2014.

\bibitem{Tse98}
D.~N.~C. Tse and S.~V. Hanly, ``{Multiaccess fading channels—part I:
  Polymatroid structure, optimal resource allocation and throughput
  capacities},'' \emph{IEEE Trans. Inf. Theory}, vol.~44, no.~7, pp.
  2796–--2815, 1998.

\bibitem{Goldsmith09}
A.~Goldsmith, S.~A. Jafar, I.~Maric, and S.~Srinivasa, ``{Breaking spectrum
  gridlock with cognitive radios: An information theoretic perspective},''
  \emph{Proc. IEEE}, vol.~97, no.~5, pp. 894–--914, 2009.

\bibitem{Neely05}
M.~J. Neely, ``{Optimal energy and delay tradeoffs for multi-user wireless
  downlinks},'' \emph{Tech. Rep. CSI-05-06-01}, vol. University of Southern
  California, 2005.

\bibitem{Boyd}
S.~Boyd and L.~Vandenberghe, \emph{Convex Optimization}.\hskip 1em plus 0.5em
  minus 0.4em\relax New York, NY: Cambridge University Press, 2004.

\bibitem{Vaart98}
A.~can~der Vaart, \emph{Asymptotic Statistics}.\hskip 1em plus 0.5em minus
  0.4em\relax Cambridge , UK: Cambridge University Press, 1998.

\bibitem{Xue13}
D.~Xue and E.~Ekici, ``{Efficient Distributed Scheduling in Cognitive Radio
  Networks in the Many-Channel Regime},'' \emph{Wi-opt}, May 2013.

\bibitem{Latreuch}
Z.~Latreuch and B.~Beladi, ``New inequalities for convexsequences with
  applications,'' \emph{Int. J. Open Problems Comput. Math.}, vol.~5, no.~3,
  pp. 15--27, Sept. 2012.

\end{thebibliography}

\appendices

{\allowdisplaybreaks
\section{Proof of Theorem \ref{thm:optimalscheduling}}
\label{proof:optimalscheduling}

The problem given in \eqref{eq:obj_linear}-\eqref{eq:sch_const} leads to the following Lagrangian relaxation:

\begin{align}
\min_{\lambda_j \geq 0, \mu \geq 0} &\max_{\boldsymbol{{\cal I}}  \in \Psi}
\E{{\cal I}_i (\boldsymbol{h},\boldsymbol{g}) R_i} +  \sum_{j\neq i}\lambda_j \left(\E{{\cal
I}_j(\boldsymbol{h},\boldsymbol{g})R_j}-\alpha_j \right) \nonumber \\
&+ \mu \left( \gamma - \sum_{l=1}^N
\E{P{\cal I}_l(\boldsymbol{h},\boldsymbol{g}) g_l}\right) \label{eq:langrange_obj}\\
%&=\min_{\lambda_j \geq  0, \mu \geq  0} \max_{\boldsymbol{{\cal I}} \in \Psi}
%\int_0^\infty \dots \int_0^\infty {\cal I}_iR_i +\sum_{j\neq
%i}\lambda_j{\cal I}_j(\boldsymbol{h},\boldsymbol{g})R_j \nonumber \\
%&-\mu P(\sum_{l=1}^N{\cal I}_l(\boldsymbol{h},\boldsymbol{g})g_l)
%p(\boldsymbol{h},\boldsymbol{g})dh_1\ldots dh_N dg_1 \ldots dg_N, \\
	\mbox{subject to } & \sum_{l=1}^N Pg_l{\cal I}_l(\boldsymbol{h},\boldsymbol{g}) \leq \nu \mbox{ and }\sum_{l=1}^N
{\cal I}_l \leq 1, \label{eq:langrange_sch_const}
\end{align}
where we ignore $\alpha_j$ and $\gamma$ in the expectation as
they are merely constants and do not affect the solution.  

Note that, since the channel rates are
independently distributed, optimizing in each fading realization results in optimal solution of statistical averages in \eqref{eq:langrange_obj}-\eqref{eq:langrange_sch_const} in dual problem. That is to say, the solution, $I_j^*=I_j^* (\boldsymbol{h},\boldsymbol{g})$
will be a stationary policy, i.e., optimizing in each fading realization results in optimal solution of the problem. For any
given values of the Lagrange multipliers, $\lambda_j$ and $\mu$, and
transmission rate vector $(R_1, \ldots, R_N)
$, the optimal policy will choose the scheduling decision in each time slot as the solution of the following Lagrangian problem:
\begin{align}
		\max_{\boldsymbol{{\cal I}}  \in \Psi} {\cal I}_i(\boldsymbol{h},\boldsymbol{g}) &\left(R_i - \mu P g_i \right) + \sum_{j\neq i} {\cal I}_j(\boldsymbol{h},\boldsymbol{g})\left(\lambda_jR_j - \mu P g_j\right) 
		\label{eq:obj_fun_sep}\\
		\mbox{subject to } & \sum_{l=1}^N Pg_l{\cal I}_l((\boldsymbol{h},\boldsymbol{g}) \leq \nu \mbox{ and } 	\sum_{l=1}^N
{\cal I}_l \leq 1.\label{eq:const_interference}	
\end{align}

Since the above optimization problem has a linear objection function and constraints, we conclude that the optimization problem lies on the boundary. Specifically, for any given
values of the Lagrange multipliers $\lambda_j$ forall $ j \neq i$ and $\mu$, the optimal policy will be deterministic policy, i.e, ${\cal I}_j(\boldsymbol{h},\boldsymbol{g}) \in \{0,1\}$. Furthermore, ${\cal I}_j(\boldsymbol{h},\boldsymbol{g}) =0$ for all $j$, i. e., the channel remains idle,  if $R_i - \mu P g_i < 0$ or $Pg_i > \nu$, and  $\lambda_jR_j - \mu P g_j < 0$ or $Pg_j > \nu$ for all $j \neq i$. 
%choose ${\cal I}_j(\boldsymbol{h},\boldsymbol{g}) = 1$ and ${\cal I}_j(\boldsymbol{h},\boldsymbol{g}) = 0$, for $j \neq i$ if the objective function is maximized and the
%constraint \eqref{eq:const_interference} is satisfied for ${\cal I}_i = 1$ and ${\cal I}_j = 0$ for $j \neq i$.
Also, for the duality gap to be zeo, the following KKT conditions should be satisfied for the optimal Lagrange multipliers \cite{Boyd}:
\begin{align}
&\E{R_i} + \sum_{j \neq i} \lambda_j^* \E{R_j} -  \sum_{l = 1}^N \mu^* P\E{g_l} = 0, \label{eq:con_obj}\\
 &\lambda_j^*\left( \E{{\cal I}_j^*R_j} - \alpha_j \right) = 0, \forall j\neq i, \label{eq:con_rate} \\
  & \mu^* \left(\gamma -\sum_{l = 1}^N \E{P{\cal}_l^* g_l}\right)  =0,  \mu^* \geq 0 \mbox{ and } \lambda_j^* \geq 0 \forall j\neq i\label{eq:con_int}
\end{align}

The conditions in \eqref{eq:con_rate} and \eqref{eq:con_int} express the fact that if a constraint is not active then its corresponding
Lagrange multiplier is 0. Since the objective function, i.e., the rate achieved by D2D pair $i$ is inversely proportional to the rates achieved by the other pairs. We conclude that $\lambda_j^* > 0$ for all $j \neq i$ for which the corresponding constraint is realized with equality.

%Also, a necessary condition for the
%existence of a feasible solution is $\alpha_j \leq \E{R_j}$, where $\E{R_j}$ is the average rate obtained by device $j$ when all channel resources are given to device $j$.

}
\section{Proof of Theorem \ref{thm:optimalcontrol}}
\label{proof:optimalcontrol}

To derive the performance bound on the control algorithm, we start with providing an upper bound on $\Delta(t)$ in the following lemma.
\begin{lemma}
\label{lemma:drift-1}

\small
\begin{align}
\Delta(t) \leq B &- \sum_{i = 1}^N \E{Q_i(t) \left({\cal I}_i(t)R_i(t) -A_i(t) \right) | \ Q_i(t)} \nonumber  \\
&-  \E{Z(t) \left( \gamma -\sum_{i = 1}^N Pg_i(t) 
\right)  | \ Z(t)}  \label{eq:delta}
\end{align}
\normalsize where $B>0$ is a constant.
\end{lemma}

\begin{IEEEproof}

By using the bounds assumed for the channel rates and the arrival rates, the following
inequalities can be obtained for each real queue:

\small
	\begin{align}
	&\E{\left(Q_i(t+1)\right)^2 - \left(Q_i(t)\right)^2 | \ Q_i(t)} \nonumber \\
	&= \E{\left(\left[ Q_i(t)-{\cal I}_i(t) R_i(t)\right]^+ + A_i(t)\right)^2- \left(Q_i(t)\right)^2 | \ Q_i(t)} \nonumber \\
	&\leq \mathbb{E}\left[Q_i(t)^2+ A_i(t)^2 + ({\cal I}_i(t)R_i(t))^2  \right. \nonumber \\ 
	&\left. -2Q_i(t)\left[ {\cal I}_i(t)R_i(t) -A_i(t)\right]-Q_i(t)^2 | \ Q_i(t)\right]  \nonumber \\
	&\leq C_1 - 2 \E{\left( {\cal I}_i(t)R_i(t) - A_i(t)\right) | \ Q_i(t)}, \label{eq:real_queue_bound}
	\end{align}
	\normalsize where $C_1 = R_{max}^2 +A_{max}^2$. The same line of derivation can be performed for the virtual queue to obtain
	
	\small
	\begin{align}
	\E{Z(t+1)^2 -Z(t)} \leq C_2 - \E{Z(t) \left( \gamma - \sum_{i = 1}^N Pg_i(t) 
\right)  | \ Z(t)} \label{eq:virtual_queue_bound},
	\end{align}
\normalsize where $C_2 = \gamma^2+n^2g_{max}^2$, and $g_{max}$ is the bound on the first moment of the interference channel gain, $g_i(t)$, i.e., $\E{g_i(t)} \leq g_{max}$, for all $i =1,\ldots,N$.

Hence, by summing \eqref{eq:real_queue_bound} over all $i = 1,\ldots, N$ and \eqref{eq:virtual_queue_bound}, and multiplying with $\frac{1}{2}$, we obtain the upper
bound on $\Delta(t)$ as given in the Lemma, where $B_1 = \frac{n\left(R_{max}^2 +A_{max}^2\right) + \gamma^2+n^2g_{max}^2}{2}$. \end{IEEEproof}

%Since the maximum transmission power is finite, in any
%interference-limited system transmission rates are bounded. Let
%$R_i^\text{max}$ be the maximum rate over uplink $i$, which depends
%on the channel states. Also assume that the arrival rate is bounded,
%i.e., $A_i^{\text{max}}$ is the maximum number of bits that may
%arrive in a block for each SU. By simple algebraic manipulations one
%can obtain a bound for the difference
%$\left(Q_i(t+1)\right)^2-\left(Q_i(t)\right)^2$ (which is
%$(R_i^\text{max})^2+(A_i^{\text{max}})^2$) and also for other queues
%to obtain the result in \eqref{eq:delta}.

%Hence, the following inequalities can be obtained for each user
%queue:

%\endproof

%Applying the above lemma, we can complete our proof. In particular,
%Lyapunov Optimization Theorem [1] suggests that a good control
%strategy is the one that minimizes the following:

%\begin{equation} \Delta^U(t)=\Delta(t) - V\E{\sum_{i=1}^N
%\left(U_i(t)\right) | \left(\mathbf{Q(t)},Z(t)\right)}
%. \label{eq:deltawithreward}
%\end{equation}
By using~\eqref{eq:delta} in the lemma, we obtain an upper bound for
\eqref{eq:deltawithreward}, as follows:

\small
\begin{align}
\Delta^U(t) & \leq  B_1  - V \E{\sum_{i=1}^N U_i(A_i(t)} - \sum_{i = 1}^N \E{Q_i(t) \left({\cal I}_i(t)R_i(t) -A_i(t) \right) | \ Q_i(t)} \nonumber  \\
& -  \E{ \gamma - Z(t) \left( \sum_{i = 1}^N Pg_i(t) 
\right)  | \ Z(t)}    \nonumber \\ 
&= B_1 - \sum_{i=1}^N  \E{\left( U_i(A_i(t) - Q_i(t)A_i(t) \right) | \ Q_i(t)} \nonumber \\
&- \sum_{i=1}^N \E{ {\cal I}_i(t) \left(Q_i(t) R_i(t) - Z(t) Pg_i(t)\right) | \ \Qv(t),Z(t)}-\gamma
\label{drift_final}
\end{align}
\normalsize

%Since $ Q_i^s(t)\left(\sum_{e \in E} \sum_{i:(i,j) \in e}
%\mu_{ij}(t) - \sum_{e \in E} \sum_{i:(j,i) \in e} \mu_{ji}(t)\right)
%= \sum_{e \in E} \sum_{i:(i,j) \in e} \mu_{ij}(t) (Q_i^s(t)-
%Q_j^s(t))$,
It is easy to observe that our proposed dynamic network control
algorithm minimizes the right hand side of \eqref{drift_final}.

If there exists a feasible region, it has been shown in \cite{Georgiadis} that
there must exist a stationary scheduling and rate control policy
that chooses the users and the arrival rates independent of queue backlogs and only with
respect to the channel statistics.  In particular, the optimal
stationary policy can be found as the solution of a deterministic
policy if the channel statistics are known a priori.

Let $U^*$ be the optimal value of the objective function of the
problem (\eqref{eq:opt-objective}-\eqref{eq:const-stability})
obtained by the aforementioned stationary policy. Also let
${x_i}^*$ be optimal traffic arrival rates found as the
solution of the same problem. In particular, the optimal input rate
${x_i}^*$ could in principle be achieved by the simple backlog-independent
admission control algorithm of including all new arrivals
$A_i(t)$ for a given pair $i$ in time $t$ independently with
probability $\zeta_i={x_i}^*/x_i$. Then, the right-hand side (RHS) of \eqref{drift_final} can be rewritten as:

\begin{align}
B_1  - VU^* - \E{Q_i(t)}\E{{\cal I}_i(t)R_i(t) -A_i(t)} \nonumber \\
-\E{Z(t)}\E{\gamma - {\cal I}_i(t) Pg_i(t)}
\label{drift_final_randomized}
\end{align}

Also, since ${x_i}^*$ is in the achievable rate region, i.e.,
arrival rates are strictly interior of the rate region, there must
exist a stationary scheduling and rate allocation policy that is
independent of queue backlogs and satisfies the following:

\begin{align}
\E{{\cal I}_i(t)R_i(t)}&\geq {x_i}^*+\epsilon_1 \mbox{ and
}\sum_{i=1}^N \E{{\cal I}_i(t)Pg_i(t)} \leq \gamma +\epsilon_2  \label{eq:opt-conds-3}
\end{align}
\normalsize

Note that as we consider stationary and ergodic policies, long-term
averages \eqref{eq:opt-conds-3} correspond to expectations of the
same variables as in \eqref{drift_final}. Clearly, any stationary
policy should satisfy \eqref{drift_final}. Recall that our proposed
policy minimizes RHS of \eqref{drift_final},
and hence, any other stationary policy (including the optimal
policy) has a higher RHS value than the one attained by our policy.
In particular, the stationary policy that satisfies
\eqref{eq:opt-conds-3}, and implements aforementioned probabilistic
admission control can be used to obtain an upper bound for the RHS
of our proposed policy. Inserting \eqref{eq:opt-conds-3} into
\eqref{drift_final_randomized}, we obtain the following upper bound for our
policy:

\begin{align}
RHS \leq B_1-\sum_{i=1}^N
\epsilon_1\mathbb{E}[Q_i(t)]
-\epsilon_2\mathbb{E}[Z(t)]-VU^*. \nonumber
\end{align}
\normalsize

This is exactly in the form of Lyapunov Optimization Theorem given
in~\cite{Georgiadis}, and hence, we can obtain bounds on the performance of the
proposed policy and the sizes of queue backlogs as given in Theorem
\ref{thm:optimalcontrol}.

{\allowdisplaybreaks\section{Proof of Lemma \ref{lemma:bound}}
\label{proof:bound_weight}

Our aim here is to obtain a bound on the average weight achieved by CADS comparing with the weight achieved by the centralized algorithm given in Section \ref{control}, i.e., the max weight algorithm. Also, we define the maximum weight, $W^*(t) = \argmax_i W_i(t) {\cal I}_i(t)$. To characterize the bound, we first introduce the following probabilities:

\vspace{-0.15in}
\small
\begin{align*}
	\Prob{\Theta(W^*(t)) = k} = \left( \prod_{n \leq k} \left[1 - \frac{1}{M-n+1}\right]^N \right) \left(1 - \left[ 1- \frac{1}{M-k+1} \right]^{N} \right). %\label{eq:prob_max_weight}.
\end{align*}
\begin{align*}
	&\Prob{\sum_{i=1}^N{\cal I}_i(t) = 1 , \Theta(W^*(t)) = k} = \left( \prod_{n \leq k} \left[1 - \frac{1}{M-n+1}\right]^N \right)  \nonumber \\
	& \ \ \ \ \ \ \ \ \ \ \ \ \ \ \ \ \ \ \ \ \ \ \ \ \ \ \ \ \ \left( {N\choose 1} \frac{1}{M-k+1}\left[ 1- \frac{1}{M-k+1} \right]^{N-1}\right). %\label{eq:prob_max_weight_suc}
\end{align*}
		
\normalsize Note that $\sum_{i=1}^N{\cal I}_i(t) = 1$	states that the contention phase takes place without a collision. Then, by using the above probabilities, we obtain the probability that the contention phase in CADS is successful given that the pair with the maximum weight contends in k$th$ minislot as: 	

\vspace{-0.15in}	
\small
\begin{align}
	&\Prob{\sum_{i=1}^N{\cal I}_i(t) = 1 \big\vert \Theta(W^*(t)) = k} = \frac{\Prob{\sum_{i=1}^N{\cal I}_i(t) = 1 , \Theta(W^*(t)) = k}}{\Prob{\Theta(W^*(t)) = k}} \nonumber \\
	& \ \ \ \ \ \ \ \ \ \ \ \ \ \ \ \ \ \ = \frac{N(M-k)^{N-1}}{(M-k+1)^N - (M-k)^N}.
	\label{eq:ineq_suc_trans}
\end{align}
		
\normalsize

The collisions in first mini-losts result in higher performance loss compared to collisions in later mini-slots. Hence, the probability in \eqref{eq:ineq_suc_trans} lets us to analyze the performance loss due to collisions in each mini-slot seperately.  

\begin{lemma}
  Let  $P_k= \Prob{\sum_{i=1}^N{\cal I}_i(t) = 1  \big\vert \Theta(W^*(t)) = k}$. Then, the probability vector, $\boldsymbol{P} = \left(P_1, P_2, \ldots, P_M \right)$, is a monotonically descreasing convex sequence.
	\label{lem:convex_seq}
\end{lemma}

\begin{IEEEproof}
We omit the proof, since it is straightforward. It is enough to show that the first and second derivative of $P_k$ with respect to $k$ are negative.
\end{IEEEproof}

Then, we bound the expected weight as:

\vspace{-0.2in}
\small
\begin{align}
&\E{\sum_{i=1}^N {\cal I}_i(t)W_i(t)} = \E{(1-M\tau)(1-\beta) W^*(t)\sum_{i=1}^N {\cal I}_i(t)} \label{eq:loss_imperfect} \\
%& = (1-M\tau)(1-\beta)\sum_{m=1}^M\E{W^*(t)\sum_{i=1}^N{\cal I}_i(t)\big\vert \Theta(W^*(t)) = m}\Prob{\Theta(W^*(t)) = m} \nonumber \\
& = (1-M\tau)(1-\beta)\sum_{k=1}^M\E{W^*(t)\big\vert \sum_{i=1}^N{\cal I}_i(t) = 1 ,\Theta(W^*(t)) = k} \nonumber \\
& \ \ \ \ \ \ \ \ \Prob{\Theta(W^*(t)) = k, \sum_{i=1}^N{\cal I}_i(t) = 1} \nonumber \\
& = (1-M\tau)(1-\beta)\sum_{k=1}^M\E{W^*(t)\big\vert \sum_{i=1}^N{\cal I}_i(t) = 1 ,\Theta(W^*(t)) = k}\nonumber\\
&\ \ \ \ \ \ \ \ \Prob{\Theta(W^*(t)) = k}\Prob{ \sum_{i=1}^N{\cal I}_i(t) = 1 \big\vert \Theta(W^*(t)) = k} \nonumber \\
& \leq  \left(\sum_{k=1}^M\E{W^*(t)\big\vert \sum_{i=1}^N{\cal I}_i(t) = 1 ,\Theta(W^*(t)) = k} \Prob{\Theta(W^*(t)) = k} \right)  \nonumber\\
&\left(\sum_{k=1}^M \frac{1}{M}\Prob{ \sum_{i=1}^N{\cal I}_i(t) = 1 \vert \Theta(W^*(t)) = k} \right) (1-M\tau)(1-\beta) \label{eq:cor_convex}\\
& = \E{W^*(t)}  \alpha \label{eq:max_weight},
\end{align} 
\normalsize
where $\alpha = (1-M\tau)(1-\beta) \left(\frac{N}{M} \sum_{k=1}^M\frac{N(M-k)^{N-1}}{(M-k+1)^N - (M-k)^N} \right)$. \eqref{eq:loss_imperfect} is by taking  into account the loss due to imperfect scheduling and mini-slot implementation. \eqref{eq:cor_convex} follows from Lemma \ref{lem:convex_seq} and  the application of corallary 1.1. in \cite{Latreuch}. Lastly, \eqref{eq:max_weight} follows from the following equality: \\

\footnotesize
$\E{W^*(t)} =	\sum_{k=1}^M\E{W^*(t)\big\vert \sum_{i=1}^N{\cal I}_i(t) = 1 ,\Theta(W^*(t)) = k}\Prob{\Theta(W^*(t)) = k}	$. 

\normalsize 
%This concludes our proof.

%Note that, in \eqref{eq:loss_imperfect}, we take into account the loss due to imperfect scheduling and mini-slot implementation. By inserting the inequality in \eqref{eq:ineq_suc_trans} into \eqref{eq:max_weight}, we obtain

%\small
%\begin{align*}
%\E{{\cal I}_i(t)W_i(t)} &\geq (1-M\tau)(1-\beta) \sum_{m=1}^M\E{W^*(t)\big\vert {\cal I}_i(t) = 1 ,\Theta(W^*(t)) = m} \\
%&\ \ \ \ \ \ \ \ \Prob{\Theta(W^*(t)) = m} \frac{N}{M}\left(1-\frac{1}{M}\right)^{N-1} \\
%&=(1-M\tau)(1-\beta)\E{W^*(t)}\frac{N}{M}\left(1-\frac{1}{M}\right)^{N-1},
%\end{align*}
%where \\
%\footnotesize
%$\E{W^*(t)} =	\sum_{m=1}^M\E{W^*(t)\big\vert {\cal I}_i(t) = 1 ,\Theta(W^*(t)) = m}\Prob{\Theta(W^*(t)) = m}	$. \normalsize 

%This concludes our proof.

\section{Proof of Theorem \ref{thm:optimalcontrol_dist_sel}}
\label{proof:interference_regulated_sel}
	The proof follows the similar lines of the proof of Theorem \ref{thm:optimalcontrol}. Thus, we skip the first lines of the proof and write the one-step expected Lyapunov drift as:
	\begin{align}
	 \Delta(t) \leq B_2 &- \sum_{i = 1}^N \E{Q_i(t) \left(A_i(t) - {\cal I}_i(t)R_i(t) \right) | \ Q_i(t)} \nonumber  \\
&-  \E{Z(t) \left( \sum_{i = 1}^N P g_i(t) -\gamma
\right)  | \ Z(t)},
\label{eq:delta_selective}
	\end{align}
	where $B_2>0$ is constant, and by using we obtain the following drift function with reward:
	
	\begin{equation} \Delta^U(t)=\Delta(t) - V\E{\sum_{i=1}^N
\left(U_i(t)\right) | \left(\mathbf{Q(t)},Z(t)\right)}
. \label{eq:deltawithreward_selective}
\end{equation}
By using~\eqref{eq:delta_selective} in the lemma, we obtain an upper bound for
\eqref{eq:deltawithreward_selective}, as follows:

\small
\begin{align}
\Delta^U(t) & <  B_2  - V \E{\sum_{i=1}^N U_i(A_i(t)} -\sum_{i = 1}^N \E{Q_i(t) \left(A_i(t) - {\cal I}_i(t)R_i(t) \right) |  Q_i(t)} \nonumber  \\
& -  \E{Z(t) \left( \sum_{i = 1}^N Pg_i(t) -\gamma
\right)  | \ Z(t)}   \nonumber  \\
&= B_2 - V \E{\sum_{i=1}^N U_i(A_i(t)}- \E{\sum_{i=1}^N I_i(t)W_i(t) | \left(\mathbf{Q(t)},Z(t)\right)} \nonumber \\
&+ \E{\sum_{i=1}^N Q_iA_i(t) | \mathbf{Q(t)} } + \E{\sum_{i=1}^N Z(t)\gamma | Z(t)}
\label{drift_selective}
\end{align}
\normalsize

By using the bound in the Lemma \ref{lemma:bound}, we obtain further bound for \eqref{drift_selective}, as follows:

\begin{align}
	\Delta^U(t) & <  B_2 - V \E{\sum_{i=1}^N U_i(A_i(t))}- \alpha \E{W^*(t)| \left(\mathbf{Q(t)},Z(t)\right)} \nonumber \\
&+ \E{\sum_{i=1}^N Q_iA_i(t) | \mathbf{Q(t)} } + \E{\sum_{i=1}^N Z(t)\gamma | Z(t)},
\label{drift_final_selective}
\end{align}	
where $\alpha = (1-M\tau)\left(\frac{N}{M}\left(1-\frac{1}{M}\right)^{N-1}\right)$. The result in \eqref{drift_final} suggests that the algorithm at least achieves $\alpha$ portion of minimized drift obtain by centralized algorithm. Again assume that $x_i^*$ denotes optimal traffic arrival rates, which is in the achievable rate region. Then, since $\alpha x^*$ is  in the $\alpha$ fraction of the achievable rate region,\footnote{In the previuos section it is defined as $\alpha\Gamma$} there must exists a stationary scheduling and rate allocation policy that is independent of queue backlogs and satisfies the following:

\begin{align}
\E{{\cal I}_i(t)R_i(t)}&\geq \alpha{x_i}^*+\epsilon_3 \mbox{ and
}\sum_{i=1}^N \E{Pg_i(t)} \leq \gamma +\epsilon_3  \label{eq:opt-conds-selec}
\end{align}
\normalsize

Note that as we consider stationary and ergodic policies, long-term
averages \eqref{eq:opt-conds-selec} correspond to expectations of the
same variables as in \eqref{drift_final}. Clearly, any stationary
policy in the $\alpha$ fraction of the achievable rate region should satisfy \eqref{drift_final}. Recall that our proposed
policy obtains $\alpha$ fraction of the minimum of the right hand side (RHS) of \eqref{drift_final},
and hence, any other stationary policy in the $\alpha$ fraction of the achievable rate region has a higher RHS value than the one attained by our policy.
In particular, the stationary policy that satisfies
\eqref{eq:opt-conds-selec}, and implements aforementioned probabilistic
admission control can be used to obtain an upper bound for the RHS
of our proposed policy. Inserting \eqref{eq:opt-conds-selec} into
\eqref{drift_final}, we obtain the following upper bound for our
policy:

\begin{align}
RHS<B_2-\sum_{i=1}^N
\epsilon_3\mathbb{E}[Q_i(t)]
-\epsilon_4\mathbb{E}[Z(t)]-V\sum_{i=1}^N U_i(\alpha x_i^*).
\label{eq:drift_opt_CSMA}
\end{align}
\normalsize

From \eqref{eq:drift_opt_CSMA}, by taking the time-average over $t = 0,1, \ldots, T-1$ and taking the limsup of $T$, we can prove Theorem \ref{thm:optimalcontrol_dist_sel}.
}

\end{document}